\documentclass[unnumsec,webpdf,contemporary,large]{oup-authoring-template}%

\graphicspath{{Fig/}}

\usepackage{amsmath,amssymb,amsfonts}
\usepackage{longtable}
\usepackage{hyperref}
\usepackage{xurl}
 \usepackage{array}
\usepackage{ragged2e}
\usepackage{booktabs}
\usepackage{xurl}
\usepackage{graphicx}
\usepackage{booktabs}
\usepackage{array}
\usepackage{ragged2e}
\usepackage{xurl}
\usepackage{float}      
\usepackage{placeins}   
\usepackage{booktabs}
\usepackage{tabularx}
\usepackage{natbib}

\usepackage{textgreek}

\usepackage{tabularray}
\theoremstyle{thmstyleone}%
\usepackage{mdframed}
\theoremstyle{thmstyletwo}%
\theoremstyle{thmstylethree}%

\begin{document}

\journaltitle{Briefings in Bioinformatics}
\copyrightyear{2026}
\pubyear{2026}
\appnotes{Paper}

\firstpage{1}

\title[G2DR: A Genotype-First Framework for Genetics-Informed Target Prioritization and Drug Repurposing ]{G2DR: A Genotype-First Framework for Genetics-Informed Target Prioritization and Drug Repurposing}

\author[1,2]{Muhammad Muneeb}
\author[1,2,$\ast$]{David B. Ascher}

\authormark{Muneeb \textit{et al}.}

\address[1]{\orgdiv{School of Chemistry and Molecular Biology},
\orgname{The University of Queensland},
\orgaddress{\street{Queen Street}, \postcode{4067}, \state{Queensland}, \country{Australia}}}

\address[2]{\orgdiv{Computational Biology and Clinical Informatics},
\orgname{Baker Heart and Diabetes Institute},
\orgaddress{\street{Commercial Road}, \postcode{3004}, \state{Victoria}, \country{Australia}}}

\corresp[$\ast$]{Corresponding author: David B. Ascher,
Email: \href{mailto:d.ascher@uq.edu.au}{d.ascher@uq.edu.au}}




\abstract{Human genetics offers a promising route to therapeutic discovery, yet practical frameworks that translate genotype-derived signal into ranked target and drug hypotheses remain limited, particularly when matched disease transcriptomics are unavailable. Here, we present G2DR, a genotype-first computational prioritization framework that propagates inherited genetic variation through genetically predicted gene expression, multi-method gene-level testing, pathway enrichment, network context, druggability, and multi-source drug--target evidence integration to generate ranked hypotheses for downstream follow-up. In a migraine case study using 733 UK Biobank participants (53 cases and 680 controls) under stratified five-fold cross-validation, we imputed genetically regulated expression across seven transcriptome-weight resources and ranked genes using a reproducibility-aware discovery score derived exclusively from training and validation data, followed by a balanced integrated score for downstream target selection. In held-out evaluation, discovery-based prioritization generalized to unseen data, achieving a gene-level ROC-AUC of 0.775 and PR-AUC of 0.475 for recovery of test-significant genes, while retaining enrichment for curated migraine-associated biology. Mapping prioritized genes to compounds through Open Targets, DGIdb, and ChEMBL yielded candidate drug sets enriched for migraine-linked and literature-associated compounds relative to a global drug background, although recovery was strongest for broader mechanism-linked and off-label therapeutic space rather than migraine-specific approved therapies. Directionality filtering further separated broadly recovered compounds from those with stronger mechanistic compatibility. G2DR should therefore be interpreted as a modular computational prioritization framework for genetics-informed hypothesis generation and follow-up in genotype-first settings, not as a clinically actionable target-identification or drug-recommendation system. All prioritized genes and compounds require independent experimental, pharmacological, and clinical validation. Supplementary data are available at \textit{Briefings in Bioinformatics} online.}


\maketitle 

\section{Introduction}
Drug repurposing remains an attractive strategy for accelerating therapeutic development because existing compounds benefit from prior pharmacological, toxicological, and manufacturing knowledge (\cite{Ashburn2004,Pushpakom2019,Nosengo2016}). Over the past two decades, computational repurposing has expanded from expression-signature matching (\cite{Lamb2006}) and network medicine (\cite{Hopkins2008,Barabasi2011,Martinez2015DrugNet}) to machine-learning and knowledge-graph approaches that integrate heterogeneous biomedical evidence at scale (\cite{Himmelstein2017,Zitnik2018}). Yet many current pipelines depend on disease-specific transcriptomic profiles (\cite{Subramanian2017,Koleti2018}), curated disease modules (\cite{Barabasi2011,Menche2015}), or historical drug--disease labels (\cite{Pushpakom2019,Himmelstein2017}), which can be limiting for newly studied phenotypes or for biobank settings where genotype and phenotype labels are available but disease-relevant molecular profiling is sparse or absent (\cite{Sudlow2015,Bycroft2018,Pushpakom2019,Wainberg2019}).

Human genetics provides a particularly attractive starting point for therapeutic prioritization because inherited variation is stable, scalable, and increasingly well characterized across large cohorts (\cite{Sudlow2015,Bycroft2018}). When measured expression is unavailable, genotype can be propagated through transcriptome imputation models to estimate genetically regulated expression (\cite{Gamazon2015,gusev2016integrative}), offering an interpretable intermediate layer between variant-level signal and candidate genes (\cite{Wainberg2019}). This strategy is appealing, but it also comes with important caveats: TWAS-style approaches prioritize genes associated with trait-linked regulatory signal, not necessarily causal effector genes (\cite{Wainberg2019}), and inference remains sensitive to linkage disequilibrium, co-regulation, tissue context, ancestry, and model architecture (\cite{Wainberg2019,Barbeira2018}). These limitations argue for integrative frameworks that improve robustness and interpretability rather than relying on any single transcriptome-prediction resource or analytical test (\cite{Wainberg2019}).

A further motivation comes from therapeutic translation. Multiple studies have shown that targets supported by human genetics are more likely to succeed in clinical development (\cite{Nelson2015,King2019}), making genetics-informed target selection an increasingly important component of drug discovery. However, the key translational challenge is not simply to detect associated genes, but to prioritize those with the strongest convergent support and then connect them to tractable pharmacological hypotheses in a transparent and reproducible manner. 

\begin{figure*}[!ht]
\centering
\includegraphics[width=0.9\textwidth]{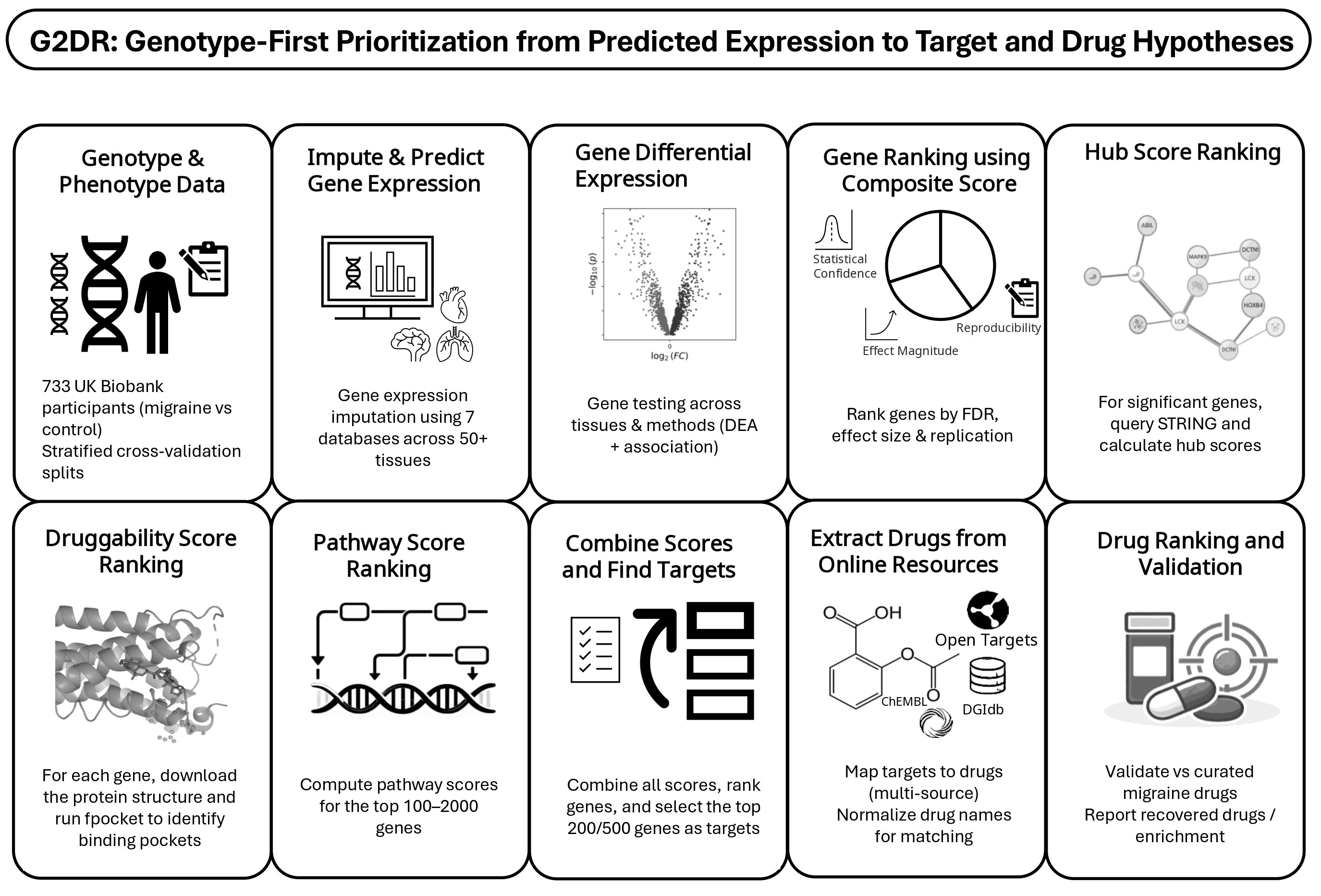}
\caption{\textbf{Overview of the G2DR framework.}
Genotype and phenotype data from the migraine cohort were partitioned by stratified cross-validation and propagated through genotype-based transcriptome imputation across multiple expression-weight resources. Covariate-adjusted predicted expression values were tested using multiple differential-expression and association methods, and significant results from the training and validation splits were aggregated into a discovery set. Genes were then ranked using a composite score that integrated reproducibility, effect magnitude, and statistical confidence, followed by pathway, network, and druggability annotation to generate an integrated target-prioritization score. Top-ranked genes were mapped to candidate compounds through Open Targets, DGIdb, and ChEMBL, and the resulting drug lists were evaluated against curated migraine-associated drug references.}
\label{abstract}
\end{figure*}

Here, we present G2DR (Figure~\ref{abstract}), a genotype-first computational framework for genetics-informed target prioritization and drug repurposing hypothesis generation. G2DR integrates genetically predicted expression (Table~\ref{tab:models}) from multiple transcriptome-weight resources (\cite{Gamazon2015,gusev2016integrative,Wainberg2019}), multi-method gene-level testing, pathway enrichment, network context, structure- and knowledge-based druggability, and multi-source drug--target evidence (\cite{Ochoa2021,Freshour2021,Gaulton2011,Gaulton2016,Wishart2018}) to generate ranked computational hypotheses for candidate targets and associated compounds, without requiring measured disease transcriptomics or supervised training on curated indication labels (\cite{Pushpakom2019}). 

\begin{table*}[!ht]
\centering
\caption{Expression weight models used for genotype-based gene expression prediction.}
\label{tab:models}
\small
\setlength{\tabcolsep}{3pt}
\renewcommand{\arraystretch}{1.15}
\resizebox{\textwidth}{!}{%
\begin{tabular}{
p{2.2cm}
p{4.6cm}
>{\RaggedRight\arraybackslash}p{9.8cm}
p{2.3cm}
}
\toprule
\textbf{Model} & \textbf{Source} & \textbf{Description} & \textbf{Gene--Tissue Pairs} \\
\midrule
MASHR &
\url{https://zenodo.org/records/7551845} &
Multivariate adaptive shrinkage models for cross-population transcriptome prediction, designed to improve TWAS performance in underrepresented populations \cite{araujo2023} &
686{,}241 (49 tissues) \\
JTI &
\url{https://zenodo.org/records/3842289} &
Joint-tissue imputation models that borrow information across tissues via shared genetic regulation, enabling both expression prediction and downstream causal inference \cite{zhou2020unified} &
533{,}141 (49 tissues) \\
CTIMP &
\url{https://github.com/yiminghu/CTIMP} &
Cross-tissue gene-expression imputation models trained with sparse group-LASSO to borrow information across GTEx tissues \cite{CTIMP} &
404{,}450 (49 tissues) \\
UTMOST &
\url{https://zenodo.org/records/3842289} &
Cross-tissue transcriptome prediction models from the UTMOST framework, hosted alongside JTI models; treated as a separate model set in this study \cite{CTIMP} &
340{,}104 (49 tissues) \\
EpiXcan &
\url{https://www.synapse.org/Synapse:syn52745052} &
Transcriptome imputation incorporating epigenomic annotations to weight SNP priors, improving prediction accuracy over standard elastic net \cite{zhang2019integrative} &
143{,}609 (14 tissues) \\
FUSION &
\url{http://gusevlab.org/projects/fusion/} &
Bayesian sparse linear mixed model and penalized regression framework for transcriptome-wide association studies \cite{gusev2016integrative} &
265{,}052 (48 tissues) \\
TIGAR &
\url{https://github.com/yanglab-emory/TIGAR} &
Nonparametric Bayesian gene expression imputation tool supporting individual- and summary-level GWAS data for transcriptome-wide association studies \cite{nagpal2019tigar} &
181{,}605 (49 tissues) \\
\bottomrule
\end{tabular}%
}
\end{table*}

We demonstrate the framework in migraine using UK Biobank data and evaluate performance at multiple levels, including held-out gene replication, enrichment for curated migraine biology, and recovery of migraine-linked drugs from prioritized targets. Framed explicitly as a computational prioritization engine for downstream experimental follow-up rather than a clinically actionable target-identification system, G2DR is best viewed as a modular proof-of-concept framework with clear room for further maturation.

\section{Methods}
\subsection{Study Design and Cohort Assembly}
A total of 733 participants (53 cases and 680 controls) who reported migraine and recorded comorbid conditions, including hypertension, asthma, depression, osteoarthritis, hypercholesterolemia, irritable bowel syndrome, hypothyroidism, hay fever, and gastroesophageal reflux disease, were extracted from the UK Biobank (\cite{Sudlow2015,Bycroft2018}). These comorbidities are commonly reported to co-occur with migraine (\cite{Buse2013Comorbidity,Bigal2009Comorbidity,Gazerani2015GutBrain,Katsarava2018Migraine}). Genotype--phenotype data were partitioned using stratified five-fold cross-validation, with each fold split into training (80\%), validation (10\%), and test (10\%) subsets to preserve the case--control ratio. Genotype quality control was applied within each training fold and propagated to validation and test subsets, including per-variant filters for minor allele frequency (MAF) $\geq 0.01$, Hardy--Weinberg equilibrium (HWE) $p \geq 1\times10^{-6}$, and variant missingness $\leq 0.1$, along with per-individual missingness $\leq 0.1$ and removal of related individuals using PLINK (\texttt{--rel-cutoff 0.125}) (\cite{Purcell2007PLINK,Anderson2010QC}).

\subsection{Genetically Predicted Gene Expression}
Genetically predicted gene expression was computed for each individual within each fold using PrediXcan-style genotype-based transcriptome imputation (\cite{Gamazon2015,Barbeira2018}), in which expression for gene $g$ in tissue $t$ is estimated as a weighted sum of SNP dosages using pre-trained expression-weight models. To maximize gene coverage and capture diverse regulatory architectures, we incorporated seven publicly available expression-weight databases trained under different statistical and biological assumptions: MASHR, JTI, CTIMP, UTMOST, EpiXcan, FUSION, and TIGAR (Table~\ref{tab:models}; full mathematical details are provided in Supplementary Methods S1). Predicted expression values were adjusted for sex and the top 10 genetic principal components using ordinary least squares regression applied to the training split only, with fitted parameters applied to validation and test splits.

\subsection{Differential Expression Analysis}
Differential expression analysis (DEA) was performed on covariate-adjusted imputed expression values for each gene, tissue, database, fold, and dataset split using eight statistical methods: LIMMA empirical Bayes moderated $t$-test (\cite{ritchie2015limma}), Welch's unequal-variance $t$-test, OLS regression, Wilcoxon rank-sum test, phenotype-label permutation test ($B = 1{,}000$ permutations), weighted logistic regression, bias-reduced Firth logistic regression, and Bayesian logistic approximation. For differential-expression methods, significance required FDR $< 0.1$ and $|\log_2\mathrm{FC}| \geq 0.5$; for association-style methods, significance required FDR $< 0.1$ and $|\mathrm{Effect}| \geq 0.5$. Complete model specifications are provided in Supplementary Methods S1.

\subsection{Gene Prioritization}
Significant genes were aggregated across all databases, tissues, folds, and methods. The discovery set was defined as $G_{\mathrm{discovery}} = G_{\mathrm{train}} \cup G_{\mathrm{val}}$, reserving $G_{\mathrm{test}}$ exclusively for out-of-sample evaluation. Each gene in $G_{\mathrm{discovery}}$ was scored using a composite importance score $S_g$ that integrates three components: reproducibility (40\%), effect magnitude (30\%), and statistical confidence (30\%). Reproducibility captures both the total number of significant hits and the breadth of support across database--tissue--method combinations. Effect magnitude is derived from method-standardized absolute effect sizes. Statistical confidence is derived from BH-adjusted FDR values. Full scoring formulas are provided in Supplementary Methods S2. Rankings were robust to alternative weights (Spearman $\rho = 0.992$; Top-100 overlap $\geq 75\%$; full stability results in Supplementary Table~S1).

Generalizability was assessed using the held-out test split as the replication target. For each gene $g$ in the tested universe $U$, a binary label $y_g = \mathbb{I}(g \in G_{\mathrm{test}})$ was defined, and $S_g$ was used as the ranking predictor. Performance was quantified using ROC-AUC, PR-AUC, and hypergeometric enrichment tests at multiple top-$K$ cutoffs. Enrichment for a curated set of 3{,}190 migraine-associated genes (\cite{Wainberg2019}) assembled via \texttt{PhenotypeToGeneDownloaderR} (\url{https://github.com/MuhammadMuneeb007/PhenotypeToGeneDownloaderR}) was assessed using the same hypergeometric framework. Full test derivations are provided in Supplementary Methods S3.

To obtain an integrated target-prioritization ranking, $S_g$ was combined with three additional evidence layers: pathway support (\textit{PathwayScore}) derived from GO, KEGG, Reactome, and Disease Ontology enrichment analyses (\texttt{clusterProfiler}, \texttt{ReactomePA}, \texttt{DOSE}); network hub score from STRING protein--protein interaction analysis; and structure- or knowledge-based druggability from DGIdb, ChEMBL, and fpocket. The integrated score was computed as $\mathrm{CoreScore}_g = 0.45 \cdot \mathrm{DE}^{\mathrm{norm}}_g + 0.25 \cdot \mathrm{Path}^{\mathrm{norm}}_g + 0.25 \cdot \mathrm{Drug}^{\mathrm{norm}}_g + 0.05 \cdot \mathrm{Hub}^{\mathrm{norm}}_g$, with DE weighted highest because it provides the primary TWAS-derived genetic link between genotype and disease. Full pathway score propagation formulas and weight justification are provided in Supplementary Methods S2.

\subsection{Drug Repurposing}
A curated migraine drug reference set of 4{,}824 normalized drug entries was assembled using \texttt{DownloadDrugsRelatedToDiseases} (\url{https://github.com/MuhammadMuneeb007/DownloadDrugsRelatedToDiseases}), which aggregates drug--disease associations from Open Targets, DrugBank, CTD, and literature-mining sources. All enrichment claims are framed relative to a global background of 139{,}597 drugs. Candidate compounds were mapped from the top $N$ ranked genes using drug--target evidence from Open Targets, DGIdb, and ChEMBL, with drug names harmonized using conservative text normalization. Each gene--drug evidence row was assigned an evidence weight based on clinical development stage, and a \textit{DrugScore} was computed by aggregating $\mathrm{GeneWeight} \times \mathrm{EvidenceWeight}$ contributions across all supporting genes.

The curated migraine drug reference set was partitioned into four development-stage tiers: Tier~1 (migraine-specific approved therapies), Tier~2 (guideline-supported acute or preventive therapies), Tier~3 (established off-label therapies), and Tier~4 (broader literature-linked compounds). Performance was assessed using multi-$K$ overlap evaluation under two complementary universes: an ALL-drugs global background (hypergeometric test valid) and a PREDICTED candidates-only universe (AUROC and AUPRC are the appropriate metrics; full results in Supplementary Table~S5).

\subsection{Directionality Assessment}

To assess mechanistic compatibility of prioritized gene--drug pairs, drug action annotations were extracted from locally aggregated evidence fields and supplemented via ChEMBL and DGIdb, harmonized into a reduced action vocabulary (inhibitor, antagonist, agonist, activator, modulator, unknown). Each pair was classified as directionally consistent when the drug action was compatible with the inferred disease-associated gene direction, inconsistent when opposed, or unclear when insufficient annotation was available. Full annotation and classification criteria are provided in Supplementary Methods S4.

\section{Results}
\label{results}
\subsection{Cross-database concordance of predicted gene expression}

A total of 733 participants (53 cases, 680 controls) were stratified into five cross-validation folds, preserving case--control balance across training (80\%), validation (10\%), and test (10\%) splits. Within each fold and split, we imputed genetically regulated expression from genotype dosages using PrediXcan-style models (\cite{Gamazon2015,Barbeira2018}).

\begin{figure*}[!ht]
\centering
\includegraphics[width=\textwidth]{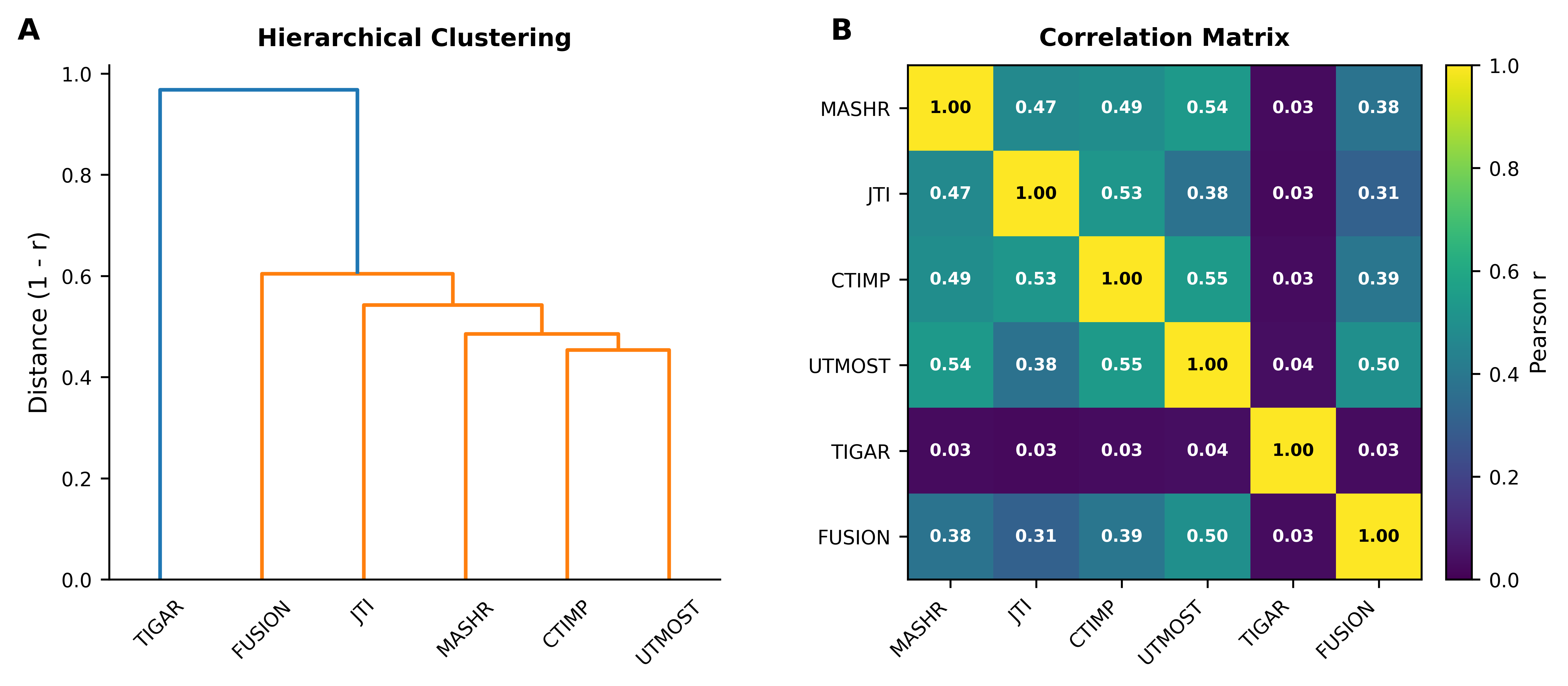}
\caption{\textbf{Cross-database concordance of predicted gene expression across expression-weight databases.}
\textbf{(A)} Hierarchical clustering of databases using distance $(1-r)$ derived from the pairwise correlation matrix.
\textbf{(B)} Pairwise correlation matrix (Pearson $r$).}
\label{fig:expr_weight_concordance}
\end{figure*}

To compare expression-weight databases, we quantified tissue-matched concordance by computing gene-wise Pearson correlations of predicted expression across individuals for genes shared by each database pair, then averaging within tissue and across folds to yield one concordance estimate per database pair and split. The highest concordance was observed for the CTIMP--UTMOST pair ($r \approx 0.55$; train $= 0.547$, validation $= 0.546$, test $= 0.546$), followed by JTI--CTIMP ($r \approx 0.53$; train $= 0.526$, validation $= 0.525$, test $= 0.525$), MASHR--UTMOST ($r \approx 0.54$; train $= 0.536$, validation $= 0.538$, test $= 0.537$), and MASHR--CTIMP ($r \approx 0.49$; train $= 0.491$, validation $= 0.492$, test $= 0.492$). Concordance with FUSION was moderate for UTMOST ($r \approx 0.50$; train $= 0.498$, validation $= 0.497$, test $= 0.499$) and lower for MASHR ($r \approx 0.38$; train $= 0.380$, validation $= 0.380$, test $= 0.378$) and JTI ($r \approx 0.31$; train $= 0.311$, validation $= 0.310$, test $= 0.311$). TIGAR showed near-zero concordance with all other databases ($r \approx 0.03$), indicating substantially different predicted-expression patterns (Figure~\ref{fig:expr_weight_concordance}).

Adjusting expression for sex and the top ten genetic principal components had negligible impact on cross-database correlations (overall mean $\Delta r = r_{\mathrm{fixed}} - r_{\mathrm{raw}} \approx -1 \times 10^{-4}$). However, the adjustment altered expression values modestly overall ($\mathrm{corr}(\text{raw},\text{fixed}) = 0.9696$, mean$|\Delta| = 0.00797$, RMSE $= 0.0353$), with the largest shifts observed in FUSION (mean$|\Delta| \approx 0.023$, RMSE $\approx 0.124$) and in EpiXcan (mean$|\Delta| \approx 0.011$, RMSE $\approx 0.078$), while TIGAR showed minimal absolute change (mean$|\Delta| \approx 6.3 \times 10^{-5}$). These results suggest that cross-database differences primarily reflect methodological and architectural variation among expression-weight resources, rather than confounding by sex or ancestry principal components.
 
\subsection{Gene differential expression analysis}
We performed 172{,}868{,}680 gene-level differential-expression and association tests across expression-weight databases, tissues, folds, dataset splits, and eight statistical methods. Applying the predefined significance and effect-size thresholds yielded 96{,}502 significant test records, corresponding to 11{,}451 unique significant genes within a tested universe of $U = 34{,}355$ genes. Given the modest cohort size and the highly layered testing design, these counts should be interpreted primarily as a broad discovery space rather than as a set of individually robust causal signals. Importantly, the main ranking conclusions were preserved under stricter thresholds in sensitivity analyses, indicating that the held-out prioritization signal does not depend solely on the most permissive operating point. Discovery candidates were defined without test leakage using training and validation only, producing $G_{\mathrm{discovery}} = G_{\mathrm{train}} \cup G_{\mathrm{val}}$ ($n = 9{,}305$), of which 1{,}189 overlapped a curated migraine-linked reference set and 8{,}116 remained outside that reference space.

Out-of-sample replication was assessed by treating test-significant genes ($T = 7{,}141$) as positives within $U$, where $G_{\mathrm{discovery}}$ recovered 4{,}995 test positives (recall $= 0.699$) with precision $= 0.537$, while smaller top-$K$ lists achieved higher precision at the cost of recall (e.g., top-100 precision $= 0.790$; Table~\ref{tab:confusion_bib}).

\begin{table*}[!ht]
\caption{\textbf{Held-out test replication for the discovery set and top-$K$ lists.}
Positives are test-significant genes ($T = 7{,}141$). TP/FP/FN are computed relative to $T$.}
\label{tab:confusion_bib}
\centering
\footnotesize
\begin{tabular}{l r r r r r}
\toprule
Predicted set & TP & FP & FN & Precision & Recall \\
\midrule
$G_{\mathrm{discovery}}$ (train $\cup$ val) & 4{,}995 & 4{,}310 & 2{,}146 & 0.537 & 0.699 \\
Top 50   & 39  & 11  & 7{,}102 & 0.780 & 0.005 \\
Top 100  & 79  & 21  & 7{,}062 & 0.790 & 0.011 \\
Top 200  & 156 & 44  & 6{,}985 & 0.780 & 0.022 \\
Top 500  & 347 & 153 & 6{,}794 & 0.694 & 0.049 \\
Top 1000 & 664 & 336 & 6{,}477 & 0.664 & 0.093 \\
\bottomrule
\end{tabular}
\end{table*}
We next evaluated the continuous discovery score as a genome-wide ranking model over the full tested universe $U$, using test-significant genes ($T = 7{,}141$) as the replication target. Under this ranking framework, the score achieved ROC-AUC $= 0.7753$ and PR-AUC $= 0.4754$, indicating meaningful separation of test-replicating genes from non-replicating genes. Given the baseline prevalence of positives in the tested universe ($T/U = 0.2079$), the observed PR-AUC corresponds to a substantial lift over random ranking. We therefore interpret the primary signal of G2DR at this stage not as binary hit identification, but as the ability to order genes such that held-out positives are preferentially concentrated toward the top of the ranked list. This distinction is important in small, class-imbalanced settings, where prioritization quality is more informative than exact gene-list identity (Table~\ref{tab:combined_enrichment}, left panel).

As an orthogonal validation, we tested enrichment for curated migraine genes ($M = 3{,}190$) within $U$. $G_{\mathrm{discovery}}$ contained 1{,}189 known genes (expected 864.01), corresponding to 1.38-fold enrichment ($p = 5.47 \times 10^{-40}$), indicating that the discovery set preferentially captures established migraine biology while retaining a large novel component (Table~\ref{tab:combined_enrichment}, right panel).

\begin{table*}[!ht]
\centering
\caption{\textbf{Hypergeometric enrichment of predicted sets for held-out test positives and curated migraine genes.} For test positives: $k_{\mathrm{exp}} = N_{\mathrm{pred}} \cdot (T/U)$, $T = 7{,}141$, $U = 34{,}355$. For migraine genes: $k_{\mathrm{exp}} = N_{\mathrm{pred}} \cdot (|M|/U)$, $|M \cap U| = 3{,}190$. FE\,=\,$k_{\mathrm{obs}}/k_{\mathrm{exp}}$.}
\label{tab:combined_enrichment}
\footnotesize
\setlength{\tabcolsep}{4pt}
\renewcommand{\arraystretch}{1.1}
\resizebox{\textwidth}{!}{%
\begin{tabular}{l r r r r l r r r r l}
\toprule
& \multicolumn{5}{c}{\textbf{Test positives ($T = 7{,}141$)}} & \multicolumn{5}{c}{\textbf{Curated migraine genes ($|M| = 3{,}190$)}} \\
\cmidrule(lr){2-6}\cmidrule(lr){7-11}
\textbf{Set} & $N_{\mathrm{pred}}$ & $k_{\mathrm{obs}}$ & $k_{\mathrm{exp}}$ & \textbf{FE} & $p$\textbf{-value} & $N_{\mathrm{pred}}$ & $k_{\mathrm{obs}}$ & $k_{\mathrm{exp}}$ & \textbf{FE} & $p$\textbf{-value} \\
\midrule
$G_{\mathrm{discovery}}$ (train $\cup$ val) & 9{,}305 & 4{,}995 & 1{,}934.13 & 2.58 & $<10^{-300}$ & 9{,}305 & 1{,}189 & 864.01 & 1.38 & $5.47\times10^{-40}$ \\
Top 50   & 50      & 39  & 10.39  & 3.75 & $7.14\times10^{-18}$  & 50      & 12  & 4.64  & 2.58 & $1.71\times10^{-3}$ \\
Top 100  & 100     & 79  & 20.79  & 3.80 & $1.55\times10^{-35}$  & 100     & 21  & 9.29  & 2.26 & $2.97\times10^{-4}$ \\
Top 200  & 200     & 156 & 41.57  & 3.75 & $1.77\times10^{-67}$  & 200     & 36  & 18.57 & 1.94 & $8.72\times10^{-5}$ \\
Top 500  & 500     & 347 & 103.93 & 3.34 & $6.94\times10^{-123}$ & 500     & 74  & 46.43 & 1.59 & $4.26\times10^{-5}$ \\
Top 1000 & 1{,}000 & 664 & 207.86 & 3.19 & $2.30\times10^{-220}$ & 1{,}000 & 132 & 92.85 & 1.42 & $2.39\times10^{-5}$ \\
\bottomrule
\end{tabular}%
}
\end{table*}

Together, these results show that integrating gene-level evidence across tissues, methods, and databases yields a reproducible prioritization that generalizes to held-out test data, performs substantially better than random ranking, and can be further refined into compact candidate gene sets for downstream network and drug-mapping analyses.
\subsection{Gene prioritization was robust to stricter significance thresholds}

To assess whether the discovery-based prioritization was sensitive to the choice of significance threshold, we repeated the same workflow under stricter FDR and effect-size criteria. Relative to the primary analysis (FDR $< 0.10$, $|\log_2\mathrm{FC}| \geq 0.50$), tightening the FDR threshold to $< 0.05$ reduced the number of significant genes in training, validation, and test from 1{,}046, 9{,}107, and 7{,}141 to 958, 7{,}861, and 5{,}313, respectively, and reduced $G_{\mathrm{discovery}}$ from 9{,}305 to 8{,}135 genes. Increasing the effect-size threshold to $|\log_2\mathrm{FC}| \geq 0.75$ at FDR $< 0.10$ reduced these counts further to 263, 6{,}241, and 4{,}599 genes, yielding $G_{\mathrm{discovery}} = 6{,}303$, while the strictest setting (FDR $< 0.05$, $|\log_2\mathrm{FC}| \geq 0.75$) produced 248 training genes, 5{,}709 validation genes, and 3{,}772 test genes, with $G_{\mathrm{discovery}} = 5{,}794$.

Despite the expected reduction in $G_{\mathrm{discovery}}$ size under stricter filtering, held-out replication remained substantial across all threshold settings. The primary analysis achieved ROC-AUC $= 0.7753$ and PR-AUC $= 0.4754$. Under FDR $< 0.05$ and $|\log_2\mathrm{FC}| \geq 0.50$, performance remained above random with ROC-AUC $= 0.7314$ and PR-AUC $= 0.3412$. Under the stricter effect-size threshold of $|\log_2\mathrm{FC}| \geq 0.75$, performance was similarly retained, with ROC-AUC $= 0.7339$ and PR-AUC $= 0.3289$ for FDR $< 0.10$, and ROC-AUC $= 0.7025$ and PR-AUC $= 0.2497$ for FDR $< 0.05$. $G_{\mathrm{discovery}}$ also remained strongly enriched for held-out test positives across all settings, with fold enrichment ranging from 2.58-fold to 3.20-fold. Enrichment for curated migraine genes was likewise preserved, ranging from 1.38-fold under the primary rule to 1.41-fold under the strictest setting. Full sensitivity results are provided in Supplementary Table~S2; together, these results indicate that the main gene-prioritization findings are not driven by permissive significance filtering.

\subsection{Component-based recovery}
To identify which components contributed the most disease-relevant signal in $G_{\mathrm{discovery}}$, we tested whether each component-specific discovery set (defined using training and validation only) was enriched for curated migraine-associated genes. For each database, tissue, and method, we defined a discovery set as the unique genes that were significant at least once within that component across folds and the remaining dimensions, and quantified over-representation of database-annotated migraine-linked genes using a fold-enrichment statistic (FE $= k_{\mathrm{obs}}/k_{\mathrm{exp}}$) with empirical $p$-values from size-matched random gene-set sampling. Full results across all databases, tissues, and methods are provided in Supplementary Table~S3.

Across expression-weight databases, MASHR showed the strongest discovery enrichment (FE $= 2.30$, $p_{\mathrm{emp}} = 1\times10^{-4}$), followed by JTI (FE $= 1.86$, $p_{\mathrm{emp}} = 6.799\times10^{-3}$) and FUSION (FE $= 1.36$, $p_{\mathrm{emp}} = 1\times10^{-4}$), whereas EpiXcan, CTIMP, UTMOST, and TIGAR showed weaker and non-significant enrichment. Across tissues, discovery enrichment highlighted a reproducible set of migraine-informative contexts, led by Brain Amygdala (FE $= 1.96$, $p_{\mathrm{emp}} = 1\times10^{-4}$), Minor Salivary Gland (FE $= 1.94$, $p_{\mathrm{emp}} = 1\times10^{-4}$), and Whole Blood (FE $= 1.83$, $p_{\mathrm{emp}} = 1\times10^{-4}$), with multiple additional tissues showing significant over-representation of database-annotated migraine-linked genes. Across methods, the discovery signal was dominated by the association-style models: Weighted Logistic and Bayesian Logistic both showed consistent enrichment (FE $\approx 1.37$, $p_{\mathrm{emp}} = 1\times10^{-4}$), while Welch $t$-test identified a smaller, disease-dense set (FE $= 1.35$, $p_{\mathrm{emp}} = 6\times10^{-4}$); low-coverage methods did not show significant discovery enrichment.

\subsection{Unified gene prioritization framework}

We next evaluated a unified gene-prioritization framework by comparing individual evidence components against both the primary composite score and the multi-source integrated score. Two complementary evaluation universes were used throughout. The \textbf{full universe} comprised all $U = 34{,}355$ tested genes; genes not in the discovery set received a score of zero, so this universe measures how well a ranking separates discovery-relevant genes from the entire tested space --- the same universe used to compute the headline ROC-AUC and PR-AUC reported above. The \textbf{discovery universe} comprised only the $n = 9{,}305$ genes in $G_{\mathrm{discovery}}$ (train $\cup$ val); here non-discovery genes are absent, so this universe measures ranking quality \emph{within} the already-selected candidates --- a strictly harder question.

Two composite scores were evaluated. The \textbf{primary composite score} ($S_g$, \textit{Importance\_Score}) integrates reproducibility, effect magnitude, and statistical confidence with weights 40/30/30. The \textbf{integrated score} additionally combines $S_g$ with pathway support, druggability evidence, and network hub score (weights DE $= 0.45$, Pathway $= 0.25$, Drug $= 0.25$, Hub $= 0.05$), and is designed for downstream target selection.

Evaluated over all 34{,}355 tested genes, effect-only ranking achieved the strongest test-replication performance (ROC-AUC $= 0.790$, PR-AUC $= 0.526$), marginally exceeding the primary composite (ROC-AUC $= 0.775$, PR-AUC $= 0.475$) and the integrated score (ROC-AUC $= 0.776$, PR-AUC $= 0.472$). The elevated AUC values across most rankings in this universe reflect the large separation between discovery genes (score $> 0$) and the majority of non-discovery genes (score $= 0$), rather than solely fine-grained within-set discrimination. For curated migraine-gene enrichment at Top-200, pathway ranking recovered 66 known genes (FE $= 3.55$, precision $= 0.330$) and hub ranking recovered 64 (FE $= 3.45$, precision $= 0.320$), both substantially exceeding the primary composite (36 genes; FE $= 1.94$) and significance-only (25 genes; FE $= 1.35$) rankings.

Within $G_{\mathrm{discovery}}$ alone, effect-only ranking remained the strongest for test replication (ROC-AUC $= 0.675$, PR-AUC $= 0.663$), recovering 145 held-out test-positive genes at Top-200. The primary composite and significance-only rankings showed similar within-set replication (ROC-AUC $\approx 0.54$--$0.55$), indicating that the headline 0.775 figure is driven in substantial part by discovery--versus--non-discovery separation rather than by within-set ranking quality alone. Pathway and hub rankings showed substantially stronger enrichment for curated migraine biology: at Top-200, pathway ranking recovered 66 database-annotated migraine-linked genes (FE $= 2.58$, precision $= 0.330$) and hub ranking recovered 64 (FE $= 2.50$, precision $= 0.320$). Direct target evidence recovered 58 known genes (FE $= 2.27$, precision $= 0.290$), whereas drug-link count and druggability alone provided weaker standalone biological enrichment.

Component-level benchmarking showed that no single evidence layer was uniformly optimal across all evaluation criteria. Effect-only ranking performed best for held-out replication, whereas pathway- and hub-based rankings were stronger for enrichment of curated migraine biology. The integrated score therefore should not be interpreted as maximizing any single benchmark metric; rather, it represents a balanced operating point designed for downstream target selection, where replication support, biological coherence, and translational tractability all matter simultaneously. In this sense, the value of integration lies in controlled trade-off rather than empirical dominance on one axis alone. Full component-wise metrics across both universes are provided in Supplementary Table~S4.

To assess whether the integrated score weights were arbitrary, we evaluated 17 alternative weighting schemes spanning nine reasonable alternatives (e.g.\ DE-heavy, pathway-heavy, drug-heavy, equal, and no-hub variants) and eight extreme stress tests (single-component schemes). Across reasonable alternatives, the mean Spearman $\rho$ against the default ranking was $0.963$ (minimum $0.866$), with a mean Top-100 gene overlap of $81.6\%$; crucially, performance metrics were virtually unchanged, with DISC-universe test ROC-AUC ranging only from $0.544$ to $0.547$ and FULL-universe test ROC-AUC from $0.775$ to $0.776$ across all nine schemes (Supplementary Table~S1). Extreme single-component schemes diverged substantially (mean $\rho = 0.682$, minimum $0.292$ for hub-only), suggesting that the integrated score draws on genuine multi-source signal rather than being dominated by any single layer. These results suggest that the default weights (DE $= 0.45$, Pathway $= 0.25$, Drug $= 0.25$, Hub $= 0.05$) lie within a stable region of parameter space, with divergence occurring only under extreme parameterizations that lack biological justification in a TWAS-first pipeline.

\subsection{Significant gene comparison against Open Targets}

We compared G2DR with the Open Targets migraine disease--target resource using the same curated migraine reference set ($M = 3{,}190$ genes). In the full analysis, G2DR ranked 9{,}305 genes across the broader genotype-first search space, whereas Open Targets returned 2{,}376 migraine-associated targets. The full G2DR ranking recovered 1{,}189 curated migraine genes (37.27\%), while Open Targets recovered 1{,}228 (38.50\%), and G2DR additionally recovered 725 curated migraine reference genes absent from Open Targets entirely. When G2DR was restricted to genes represented in the Open Targets target space ($n = 823$), Top-50 precision increased from 22.00\% to 46.00\% and Top-200 precision increased from 23.00\% to 55.50\%, approaching but not matching the standalone Open Targets precision (Top-50: 92.00\%), which reflects its disease-curated, pre-filtered design rather than a direct performance comparison. Full comparison metrics are provided in Supplementary Table~S6. These comparisons are complementary rather than directly competitive: Open Targets offers higher precision within a curated space, while G2DR extends target hypothesis generation beyond existing disease annotations, recovering a substantial set of curated migraine reference genes that are absent from Open Targets entirely.

\subsection{Drug mapping and candidate enrichment}

Using the ranked migraine gene lists (top $N = 200$ and $N = 500$), we aggregated gene--drug evidence from Open Targets, DGIdb, and ChEMBL to generate ranked candidate compounds. For $N = 200$, the pipeline compiled 5{,}348 gene--drug evidence rows and yielded 3{,}963 unique predicted drugs. For $N = 500$, evidence increased to 11{,}234 rows and the predicted set expanded to 7{,}981 unique drugs.

We evaluated overlap with a curated reference set of 4{,}824 normalized migraine-associated drugs under two complementary evaluation frames. First, an \textbf{ALL-drugs universe} (139{,}597 background drugs) tests capture and enrichment: whether the returned candidates include migraine-linked drugs above chance relative to the global pharmacopeia. Second, a \textbf{PREDICTED universe} (3{,}963 or 7{,}981 returned drugs) evaluates ranking within the returned candidate set: whether reference migraine-associated drugs are preferentially placed toward the top among returned compounds. Note that FE values below 1.0 in the PREDICTED universe reflect the fact that the reference migraine drug set exceeds the returned candidate pool in size, making standard hypergeometric enrichment interpretation inapplicable in that frame; AUROC and AUPRC are the appropriate metrics for within-set ranking performance.

Under the ALL-drugs background, both gene sets showed strong enrichment of known migraine drugs. For $N = 200$, recovery increased from $5/20$ (Precision@20 $= 0.25$, FE $= 7.23$, $p = 4.93\times10^{-4}$) to $51/100$ (Precision@100 $= 0.51$, FE $= 14.76$, $p = 4.20\times10^{-47}$) and $205/500$ (Precision@500 $= 0.41$, FE $= 11.86$, $p = 6.19\times10^{-161}$). For $N = 500$, early enrichment improved further ($8/20$, Precision@20 $= 0.40$, FE $= 11.58$, $p = 1.75\times10^{-7}$; $53/100$, Precision@100 $= 0.53$, FE $= 15.34$, $p = 4.48\times10^{-50}$; $213/500$, Precision@500 $= 0.426$, FE $= 12.33$, $p = 1.67\times10^{-171}$). These results indicate that both gene inputs yield drug sets highly enriched for established migraine-relevant compounds relative to a global drug background.

Within the returned candidates, for $N = 200$, 353 curated migraine drugs were present among 3{,}963 predicted drugs and the ranking achieved AUROC $= 0.8004$ and AUPRC $= 0.3528$. For $N = 500$, 527 curated migraine drugs were present among 7{,}981 predicted drugs with AUROC $= 0.8152$ and AUPRC $= 0.3311$. Expanding the input from 200 to 500 genes increased the number of recovered reference drugs and modestly improved AUROC, while AUPRC decreased slightly due to the larger candidate universe. Together, the two-universe evaluation separates capture from ranking: the strong fold-enrichment and hypergeometric significance under the ALL universe supports disease-relevant signal in the upstream gene prioritization, while AUROC and AUPRC suggest that reference migraine-associated drugs tend to appear earlier than non-reference drugs within the returned set. Full multi-$K$ overlap results for both $N = 200$ and $N = 500$ under both universes are provided in Supplementary Table~S5.

When we mapped the predicted drug set back to disease indications in the aggregated drug--disease database, migraine was not always the top recovered indication. Instead, cardiometabolic, inflammatory, neuropsychiatric, and seizure-related categories frequently ranked highly. This is biologically plausible because migraine has a well-established comorbidity landscape and shares mechanisms, risk factors, and therapeutic overlap with multiple neurological and systemic disorders (\cite{Buse2013Comorbidity,Bigal2009Comorbidity,Gazerani2015GutBrain,Katsarava2018Migraine}). Disease-label recovery should therefore be interpreted as reflecting shared pharmacology and overlapping biology across comorbid conditions rather than as a migraine-exclusive signal. We partitioned the curated migraine drug reference set ($n = 4{,}824$ normalized drugs) into four evidence-oriented tiers: migraine-specific approved therapies (Tier~1; triptans, ditans, gepants, CGRP monoclonals, ergot derivatives), guideline-supported acute or preventive therapies (Tier~2), established off-label therapies (Tier~3), and broader literature-linked compounds (Tier~4). Recovery of each tier was evaluated within the ranked candidate drug list against a global background of 139{,}597 drugs.

Across the pooled migraine-drug reference set, the ranked list recovered migraine-linked compounds at rates above random expectation, but tiered evaluation showed that this recovery was uneven across evidence classes. No Tier~1 migraine-specific approved therapies were recovered within the Top-200 ranked drugs, whereas overlap was more evident for guideline-supported, off-label, and broader literature-linked compounds. These results indicate that the current framework is better at surfacing broader mechanism-linked and repurposing-relevant pharmacology than at rediscovering the most migraine-specific modern therapeutic classes. We therefore interpret the drug-level signal as evidence of biologically meaningful translational expansion rather than indication-specific clinical precision (Table~\ref{tab:tiered_migraine_drug_benchmark}).

\begin{table*}[ht]
\centering
\caption{\textbf{Tiered migraine-evidence benchmark of the ranked drug list.} Tier~1: migraine-specific approved therapies ($n_{\mathrm{ref}}=135$); Tier~2: guideline-supported therapies ($n_{\mathrm{ref}}=92$); Tier~3: established off-label therapies ($n_{\mathrm{ref}}=29$); Tier~4: broader literature-linked compounds ($n_{\mathrm{ref}}=4{,}568$). Fold enrichment (FE) and hypergeometric $p$-values are computed relative to a global background of 139{,}597 drugs. High FE values for Tiers~2--3 reflect low absolute counts and should be interpreted accordingly.}
\label{tab:tiered_migraine_drug_benchmark}
\small
\setlength{\tabcolsep}{5pt}
\begin{tabular}{lcccccccc}
\toprule
\textbf{Tier} & \textbf{Top-20} & \textbf{Top-50} & \textbf{Top-100} & \textbf{Prec@100} & \textbf{Rec@100} & \textbf{FE@100} & $p$\textbf{@100} & \textbf{Top-200} \\
\midrule
Tier~1: Migraine-specific approved & 0 & 0 & 0 & 0.00 & 0.000 & 0.00 & 1.00 & 0 \\
Tier~2: Guideline-supported & 0 & 1 & 2 & 0.02 & 0.022 & 30.35 & $2.04\times10^{-3}$ & 2 \\
Tier~3: Established off-label & 1 & 1 & 1 & 0.01 & 0.034 & 48.14 & $2.06\times10^{-2}$ & 1 \\
Tier~4: Broad literature-linked & 4 & 18 & 48 & 0.48 & 0.011 & 14.67 & $7.08\times10^{-44}$ & 86 \\
\bottomrule
\end{tabular}
\end{table*}

\subsection{Directionality of prioritized gene--drug pairs}

To assess mechanistic plausibility beyond drug enrichment alone, we evaluated the directionality of all gene--drug pairs derived from the top 200 genes of the final integrated migraine ranking. Across $4{,}861$ unique gene--drug pairs, 549 (11.3\%) were classified as directionally consistent, 618 (12.7\%) as inconsistent, and $3{,}694$ (76.0\%) as unclear (Table~\ref{tab:drug_directionality_summary}). When restricted to the 614 pairs involving drugs from the curated migraine reference set, 50 (8.1\%) were consistent, 83 (13.5\%) inconsistent, and 481 (78.3\%) unclear. The proportions were broadly stable across both universes, indicating that the directionality signal does not selectively concentrate in migraine-annotated compounds.

Among the 50 directionally consistent migraine-drug pairs, several biologically interpretable clusters emerged. First, aspirin--\textit{GSTP1} (drug rank 26; \textit{GSTP1} higher in cases) was classified as consistent based on a ChEMBL inhibitor annotation; this pair should be interpreted with caution because aspirin's primary mechanism is COX-1/COX-2 inhibition, and the GSTP1 annotation likely reflects indirect or in vitro effects rather than direct pharmacological inhibition. Second, the cardiac glycosides digoxin and digitoxin (drug ranks 126 and 133) were consistent against \textit{ATP1A4} (higher in cases; gene rank 2), which encodes a Na\textsuperscript{+}/K\textsuperscript{+}-ATPase $\alpha$4 subunit; cardiac glycosides are established Na\textsuperscript{+}/K\textsuperscript{+}-ATPase inhibitors, making the directionality pharmacologically coherent, though \textit{ATP1A4} itself is supported by only a single-family case report in migraine and should be considered a low-evidence candidate gene rather than an established migraine locus. Third, several GLP-1 receptor agonists (exenatide, liraglutide, semaglutide, lixisenatide; drug ranks 142--146) were mapped against \textit{ALPL} (lower in cases; gene rank 6) via DGIdb annotations; this mapping should be interpreted cautiously because GLP-1 agonists act canonically through GLP1R rather than ALPL directly, and the annotation likely reflects indirect effects on alkaline phosphatase activity rather than direct pharmacological agonism. Fourth, $\alpha_1$-adrenergic receptor antagonists including phenoxybenzamine, phentolamine, prazosin, and carvedilol (drug ranks 377--551) were consistent against \textit{ADRA1A} (higher in cases; gene rank 150), where inhibition of elevated adrenergic signalling is mechanistically plausible. Fifth, amitriptyline (drug rank 438; a guideline-supported migraine prophylactic) showed consistent directionality against \textit{AADAC} (higher in cases; gene rank 119) via DGIdb inhibitor annotations. 


Directionality filtering materially refined the candidate space by separating broadly recovered compounds from those with stronger mechanistic compatibility. Among migraine-linked or biologically plausible candidates, only a subset remained directionally consistent, while others were directionally inconsistent or unresolved because the mapped action did not align clearly with the inferred disease-associated gene direction. This distinction matters because enrichment alone can recover compounds that are pharmacologically adjacent to migraine biology without necessarily supporting a coherent repurposing hypothesis. Accordingly, we treat directionally consistent pairs as higher-priority computational hypotheses, whereas inconsistent or unresolved pairs should be viewed as context signals rather than actionable leads.
\begin{table*}[!ht]
\centering
\small
\setlength{\tabcolsep}{6pt}
\caption{\textbf{Summary of directionality classification for unique prioritized gene--drug pairs} from the final integrated migraine ranking. Results are shown for all pairs derived from the top 200 ranked genes and separately for pairs involving drugs from the curated migraine reference set ($n_{\mathrm{mig}}=614$).}
\label{tab:drug_directionality_summary}
\begin{tabular}{lccccc}
\toprule
\textbf{Directionality class} & \multicolumn{2}{c}{\textbf{All pairs ($n=4{,}861$)}} & & \multicolumn{2}{c}{\textbf{Migraine drugs only ($n=614$)}} \\
\cmidrule(lr){2-3}\cmidrule(lr){5-6}
 & \textbf{Count} & \textbf{\%} & & \textbf{Count} & \textbf{\%} \\
\midrule
Consistent   & 549       & 11.3\%  & & 50  & 8.1\%  \\
Inconsistent & 618       & 12.7\%  & & 83  & 13.5\% \\
Unclear      & $3{,}694$ & 76.0\%  & & 481 & 78.3\% \\
\midrule
Total        & $4{,}861$ & 100.0\% & & 614 & 100.0\% \\
\bottomrule
\end{tabular}
\end{table*}

\begin{table*}[!ht]
\centering
\scriptsize
\setlength{\tabcolsep}{3pt}
\renewcommand{\arraystretch}{1.05}
\caption{\textbf{Representative directionally consistent gene--drug pairs from the migraine reference set.} Pairs were classified as consistent when the known drug action was mechanistically compatible with the inferred disease-associated direction of the target gene. Drug rank is from the integrated gene-to-drug scoring pipeline; gene rank is from the \textit{Combined\_Score} integrated gene prioritization. Evidence source indicates the database providing mechanism-of-action annotation. The \textit{MRPL36} antibiotic cluster (rows 3--9) reflects DGIdb-derived mitochondrial ribosomal inhibitor annotations and should be interpreted with caution.}
\label{tab:drug_directionality_consistent}
\resizebox{\textwidth}{!}{%
\begin{tabular}{c l l c p{2.4cm} p{2.8cm} p{1.8cm} c c}
\toprule
\textbf{Drug rank} & \textbf{Drug} & \textbf{Gene} & \textbf{Gene rank} & \textbf{Disease direction} & \textbf{Drug action} & \textbf{Evidence source} & \textbf{Approved} & \textbf{Known migraine gene} \\
\midrule
 26  & Aspirin         & GSTP1  &  31 & Higher in cases & Inhibitor & ChEMBL & Yes & Yes \\
126  & Digoxin         & ATP1A4 &   2 & Higher in cases & Inhibitor & DGIdb  & Yes & Yes \\
133  & Digitoxin       & ATP1A4 &   2 & Higher in cases & Inhibitor & DGIdb  & Yes & Yes \\
142  & Exenatide       & ALPL   &   6 & Lower in cases  & Agonist   & DGIdb  & Yes & No  \\
144  & Liraglutide     & ALPL   &   6 & Lower in cases  & Agonist   & DGIdb  & Yes & No  \\
146  & Semaglutide     & ALPL   &   6 & Lower in cases  & Agonist   & DGIdb  & Yes & No  \\
 54  & Chloramphenicol & MRPL36 &  18 & Higher in cases & Inhibitor & DGIdb  & Yes & No  \\
 86  & Minocycline     & MRPL36 &  18 & Higher in cases & Inhibitor & DGIdb  & Yes & No  \\
162  & Clarithromycin  & MRPL36 &  18 & Higher in cases & Inhibitor & DGIdb  & Yes & No  \\
379  & Prazosin        & ADRA1A & 150 & Higher in cases & Inhibitor & DGIdb  & Yes & Yes \\
512  & Carvedilol      & ADRA1A & 150 & Higher in cases & Inhibitor & DGIdb  & Yes & Yes \\
438  & Amitriptyline   & AADAC  & 119 & Higher in cases & Inhibitor & DGIdb  & Yes & No  \\
257  & Empagliflozin   & SLC5A2 &  45 & Higher in cases & Inhibitor & DGIdb  & Yes & No  \\
222  & Glycine         & GRIN2B &  44 & Lower in cases  & Agonist   & DGIdb  & Yes & Yes \\
497  & Loxapine        & KCNT1  & 133 & Lower in cases  & Activator & DGIdb  & Yes & Yes \\
\bottomrule
\end{tabular}%
}
\end{table*}

\subsection{Top gene and drug validation panel}
To provide a concise summary of the final integrated prioritization, we assembled top-gene and top-drug evidence panels from the \textit{Combined\_Score}-ranked gene list and the downstream drug-mapping and directionality outputs. The top-gene panel (Table~\ref{tab:top_gene_panel}) reports the 10 highest-ranked genes together with their evidence support, inferred disease direction, migraine relevance, and linked druggability information. The top-drug panel (Table~\ref{tab:top_drug_panel}) reports the 10 highest-ranked drugs with merged target gene information, approval status, and directionality classification.

\paragraph{Top gene highlights.}
Among the top 10 integrated-score genes, \textit{ATP1A4} (rank 2; higher in cases; known migraine gene; 24 linked drugs; Consistent directionality) was the most directly disease-relevant, encoding the Na\textsuperscript{+}/K\textsuperscript{+}-ATPase $\alpha$4 subunit and linking to cardiac glycoside inhibitors including digoxin and digitoxin. \textit{ALPL} (rank 6; lower in cases; Tier~1 confidence; 96 linked drugs; Consistent) was the most druggable top gene, linked to GLP-1 agonists through DGIdb annotations. \textit{HLA-DQB1} (rank 3; lower in cases; known migraine gene; 55 linked drugs) had the broadest drug connectivity but showed inconsistent directionality. \textit{MRPL36} (rank 18; higher in cases; 67 linked drugs; Consistent) was linked to a cluster of antibiotic and antimicrobial drugs through mitochondrial ribosomal inhibition annotations. Among the 49 database-annotated migraine-linked genes present in the top 200, additional highly ranked examples included \textit{GSTP1} (rank 31; Consistent; aspirin-GSTP1 was the most evidenced approved consistent pair), \textit{GRIN2B} (rank 44; 110 linked drugs; known migraine gene), and \textit{OSGEP} (rank 58; highest composite \textit{Importance\_Score} of 0.682; known migraine gene). Several top-ranked genes --- including \textit{GRHPR}, \textit{MRPL21}, \textit{DHX15}, and \textit{GCDH} --- had few or no linked drugs.

\paragraph{Top drug highlights.}
Among the top 10 ranked drugs, lifitegrast (rank 6; approved; targets \textit{ITGB2} and \textit{ICAM1}, both higher in cases; Consistent) was the only approved drug in the top 10 with directional consistency, acting as an integrin inhibitor compatible with elevated integrin subunit expression. Oleclumab (rank 12; Phase~3; targets \textit{NT5E}; Consistent) and ezatiostat (rank 13; Phase~2; targets \textit{GSTP1}; Consistent) provided the clearest non-approved consistent candidates. Neramexane (rank 17; Phase~3; targets \textit{GRIN2B} and \textit{CHRNA10}; Consistent) was the highest-ranking investigational drug with a migraine-biology rationale through NMDA receptor modulation. Several top-ranked approved drugs showed directionally inconsistent mappings, including metformin (rank 2; Inconsistent against \textit{NDUFV1} and \textit{NDUFS6}), amantadine (rank 5; Inconsistent against \textit{GRIN2B}), dibotermin alfa (rank 4; Inconsistent against \textit{BMPR1B}), and apremilast (rank 10; Inconsistent against \textit{PDE4C}). Among the top 20 migraine-reference drugs, aspirin (rank 26; Consistent against \textit{GSTP1}) was the highest-ranking approved drug with directional support, while metformin, esketamine, and memantine appeared at ranks 2, 7, and 8 respectively, consistent with their presence in the migraine drug compendium.

\begin{table*}[!ht]
\centering
\small
\setlength{\tabcolsep}{3pt}
\caption{\textbf{Top 10 prioritized genes from the final integrated migraine ranking.} Genes are ranked by \textit{Combined\_Score} (integrated: DE + pathway + druggability + hub). KnMig = known migraine gene; Tier = confidence tier from primary composite ranking; NDrugs = number of unique linked drugs; ApprDrug = has at least one approved linked drug; DirBest = best directionality support across linked drug pairs.}
\label{tab:top_gene_panel}
\resizebox{\textwidth}{!}{%
\begin{tabular}{c l c c p{2.2cm} c p{2.0cm} c c c c c c}
\toprule
\textbf{Rank} & \textbf{Symbol} & \textbf{CombScore} & \textbf{ImpScore} & \textbf{Disease direction} & \textbf{KnMig} & \textbf{Confidence tier} & \textbf{Hits} & \textbf{Tissues} & \textbf{DBs} & \textbf{NDrugs} & \textbf{ApprDrug} & \textbf{DirBest} \\
\midrule
 1 & APOBEC3G & 0.936 & 0.435 & Higher in cases & No  & Tier3\_Exploratory & 100 & 29 & 1 & 98 & Yes & Unclear \\
 2 & ATP1A4   & 0.934 & 0.531 & Higher in cases & Yes & Tier3\_Exploratory & 18 & 3 & 3 & 24 & Yes & Consistent \\
 3 & HLA-DQB1 & 0.933 & 0.410 & Lower in cases  & Yes & Tier2\_Moderate    & 32 & 15 & 2 & 55 & Yes & Inconsistent \\
 4 & PSMB5    & 0.932 & 0.413 & Lower in cases  & No  & Tier2\_Moderate    & 40 & 13 & 2 & 33 & Yes & Inconsistent \\
 5 & SLC1A1   & 0.932 & 0.491 & Lower in cases  & No  & Tier2\_Moderate    & 103 & 15 & 2 & 54 & Yes & Inconsistent \\
 6 & ALPL     & 0.931 & 0.412 & Lower in cases  & No  & Tier1\_High        & 35 & 5 & 1 & 96 & Yes & Consistent \\
 7 & AQP5     & 0.930 & 0.419 & Lower in cases  & No  & Tier2\_Moderate    & 42 & 10 & 2 & 29 & Yes & Consistent \\
 8 & CBL      & 0.930 & 0.385 & Unclear         & No  & Tier3\_Exploratory & 4 & 2 & 1 & 29 & Yes & Unclear \\
 9 & GRHPR    & 0.929 & 0.485 & Higher in cases & No  & Tier3\_Exploratory & 206 & 16 & 1 & 2 & No  & Unclear \\
10 & MRPL21   & 0.929 & 0.447 & Higher in cases & Yes & Tier2\_Moderate    & 93 & 31 & 1 & 1 & No  & Unclear \\
\bottomrule
\end{tabular}}
\end{table*}

\begin{table*}[!ht]
\centering
\small
\setlength{\tabcolsep}{3pt}
\caption{\textbf{Top 10 prioritized drug candidates from the final integrated migraine repurposing analysis.} Drugs are ranked by \textit{DrugScore}; target genes are merged per drug. KnMig = drug present in curated migraine reference set; NTarg = number of unique target genes among top 200; Dir = best directionality classification across gene--drug pairs.}
\label{tab:top_drug_panel}
\resizebox{\textwidth}{!}{%
\begin{tabular}{c l c c c c p{3.8cm} p{2.6cm} p{2.2cm} p{2.0cm}}
\toprule
\textbf{Rank} & \textbf{Drug} & \textbf{Phase} & \textbf{KnMig} & \textbf{Approved} & \textbf{NTarg} & \textbf{Target genes} & \textbf{Disease directions} & \textbf{Directionality} & \textbf{Evidence sources} \\
\midrule
 1 & Ataluren        & APPROVED & No  & Yes & 4 & RPL28, RPS14, RPL12, RPS16 & Unclear, lower & Unclear      & DGIdb, OpenTargets \\
 2 & Metformin       & APPROVED & Yes & Yes & 2 & NDUFV1, NDUFS6             & Lower in cases & Inconsistent & OpenTargets, DGIdb \\
 3 & Voxelotor       & APPROVED & No  & Yes & 1 & HBB                        & Lower in cases & Unclear      & OpenTargets \\
 4 & Dibotermin alfa & APPROVED & No  & Yes & 1 & BMPR1B                     & Higher in cases & Inconsistent & OpenTargets \\
 5 & Amantadine      & APPROVED & No  & Yes & 1 & GRIN2B                     & Lower in cases & Inconsistent & OpenTargets \\
 6 & Lifitegrast     & APPROVED & No  & Yes & 2 & ITGB2, ICAM1               & Higher in cases & Consistent   & OpenTargets \\
 7 & Esketamine      & APPROVED & Yes & Yes & 1 & GRIN2B                     & Lower in cases & Unclear      & OpenTargets, DGIdb \\
 8 & Memantine       & APPROVED & Yes & Yes & 1 & GRIN2B                     & Lower in cases & Unclear      & OpenTargets, DGIdb \\
 9 & Midostaurin     & APPROVED & No  & Yes & 7 & CBL, BMPR1B, CDK3, LATS2, PRKD3, SEPTIN9, PEG3 & Mixed & Inconsistent & OpenTargets, DGIdb \\
10 & Apremilast      & APPROVED & No  & Yes & 1 & PDE4C                      & Lower in cases & Inconsistent & OpenTargets \\
\bottomrule
\end{tabular}}
\end{table*}

\subsection{Biological interpretation of prioritised genes and drug repurposing candidates}

The 200 genes prioritised by the integrated G2DR pipeline converged on several biological themes with established or plausible relevance to migraine pathophysiology. Of these, 49 (24.5\%) had prior evidence linking them to migraine or closely related headache phenotypes based on database annotation from OMIM, DisGeNET, and HPO, whereas the remainder represent computationally prioritized candidates without established migraine association. These novel signals should be treated as hypothesis-generating outputs of the prioritization pipeline rather than validated migraine targets, and their biological interpretation below is offered to motivate experimental follow-up rather than to make mechanistic claims. Overall, the prioritised genes mapped to pathways related to glutamatergic and ion-channel excitability, mitochondrial and metabolic function, immune and inflammatory signalling, cerebrovascular and extracellular-matrix integrity, and broader neuronal and synaptic maintenance, consistent with the multifactorial biology highlighted by large migraine genetic studies \cite{Hautakangas2022,Gormley2016}.

Among the most interpretable signals, \textit{GRIN2B} and \textit{SLC1A1} supported glutamatergic hyperexcitability and cortical-spreading-depression-related biology \cite{Gasparini2016,Crivellaro2021}, while \textit{SCN1B}, \textit{KCNT1}, \textit{KCNS2}, and \textit{ATP1A4} pointed to broader disturbance of ion-channel homeostasis and neuronal firing. A second major component comprised mitochondrial and metabolic genes, including \textit{NDUFV1}, \textit{NDUFS6}, \textit{MRPL21}, \textit{MRPL36}, \textit{TUFM}, \textit{VARS2}, \textit{FXN}, \textit{ETFB}, \textit{COASY}, and \textit{HAAO}, consistent with longstanding hypotheses that impaired cerebral bioenergetics contributes to migraine susceptibility \cite{Gross2019,Curto2016}. Immune and inflammatory signals were also prominent, including multiple HLA-region genes together with \textit{RELA}, \textit{ICAM1}, \textit{CLEC7A}, \textit{LY96}, \textit{UNC93B1}, and \textit{SERPINE1}, in keeping with evidence that neuroinflammatory and immune-trafficking mechanisms contribute to trigeminovascular sensitisation \cite{Karatas2013,Yang2014,Bernecker2022}. In parallel, neurovascular and extracellular-matrix genes such as \textit{HTRA1}, \textit{HSPG2}, \textit{GNAQ}, \textit{TIE1}, \textit{EGFL7}, \textit{LAMB1}, \textit{F12}, and multiple collagen-related genes supported a vascular-integrity component, consistent with the broader overlap between migraine and cerebral small-vessel or neurovascular disorders \cite{Hara2009,Shirley2013,Maas2018}. Together, these results suggest that the prioritised set reflects convergence across excitability, metabolism, inflammation, and neurovascular maintenance rather than a single dominant molecular pathway.

The downstream drug-mapping analysis likewise recovered a broad migraine-relevant pharmacological landscape rather than only migraine-specific therapies. Using the top 200 genes, the pipeline identified 3{,}963 unique candidate drugs from 5{,}348 gene--drug evidence rows, and 353 curated migraine-associated drugs were present within the predicted set. Enrichment against the global drug background was strong, indicating that the upstream gene prioritisation captures disease-relevant pharmacology even though the output did not capture classical migraine medications. At the top of the ranked list, several recovered migraine-linked drugs were already notable, including metformin (rank 2), esketamine (rank 7), memantine (rank 8), aspirin (rank 26), dexamethasone (rank 48), lidocaine (rank 66), zonisamide (rank 79), indomethacin (rank 85), and acetaminophen (rank 98). Additional recovered migraine-associated or migraine-relevant drugs were present lower in the ranked set, including verapamil (rank 213), valproic acid (rank 253), celecoxib (rank 340), diclofenac (rank 343), amitriptyline (rank 438), topiramate (rank 491), and ergotamine tartrate (rank 521), showing that the pipeline recovered both acute-care and preventive classes as well as broader off-label and mechanism-linked therapies \cite{Kirthi2013,Derry2013,Marmura2015,Singh2008,Colman2008,Schwenk2022,Pomeroy2017,Xu2023}.

Directionality filtering was particularly useful for separating broadly recovered drugs from those with clearer mechanistic compatibility. Among migraine-linked drugs, aspirin showed consistent directionality at \textit{GSTP1}; amitriptyline was directionally consistent at \textit{AADAC}; and a cluster of antibiotic-derived mappings including chloramphenicol, minocycline, clarithromycin, erythromycin, tigecycline, and tobramycin were directionally consistent through \textit{MRPL36}, although these latter signals should be interpreted cautiously because they arise from mitochondrial ribosomal annotations rather than direct migraine therapeutic use. Several non-reference but biologically interesting candidates were also directionally consistent as computational hypotheses, including lifitegrast at the \textit{ITGB2}--\textit{ICAM1} axis, neramexane at \textit{GRIN2B}/\textit{CHRNA10}, digoxin and digitoxin at \textit{ATP1A4}, and GLP-1 receptor agonists such as exenatide, liraglutide, and semaglutide, mapped to \textit{ALPL} via an indirect DGIdb annotation (GLP-1 agonists act canonically through GLP1R rather than ALPL directly); notably, GLP-1 agonists have shown promising signals in emerging clinical reports of chronic migraine management \cite{Semba2016,Parsons2009,Podkowa2023,Braca2025}, although all such pairs require direct experimental and clinical validation before any repurposing conclusions can be drawn. By contrast, several high-ranking recovered migraine drugs were directionally inconsistent, including metformin at \textit{NDUFV1}/\textit{NDUFS6}, dipyridamole at \textit{PDE4C}, carbamazepine, zonisamide, eslicarbazepine, oxcarbazepine, phenytoin, and lidocaine at \textit{SCN5A}, as well as verapamil in its mapped targets. Other prominent recovered drugs, including esketamine, memantine, dexamethasone, indomethacin, acetaminophen, and valproic acid, were retained but remained directionally unresolved. Overall, these results indicate that the G2DR framework recovers a substantial set of migraine-relevant compounds, but that directional filtering is necessary to distinguish broadly associated drugs from those with stronger mechanistic coherence for repurposing.
Taken together, the drug-mapping results suggest that G2DR is currently strongest as a framework for broad genetics-anchored translational hypothesis generation. It expands beyond curated disease-target space, recovers migraine-linked pharmacology above background, and benefits from directionality-aware refinement, but it does not yet recover migraine-specific approved therapies with high fidelity. This operating profile is consistent with a framework designed for prioritization and follow-up rather than definitive repurposing recommendation.

\section{Discussion}
\label{discussion}

We present G2DR as a genotype-first computational prioritization framework for settings in which genotype and phenotype labels are available but matched disease transcriptomics are limited or absent. The principal contribution is not a new TWAS algorithm or a clinically actionable target-identification system, but a modular workflow that converts inherited genetic signal into ranked target and compound hypotheses through multiple complementary evidence layers. In a migraine proof-of-concept, the framework generalized to held-out data at the gene-prioritization stage, recovered established migraine biology, and produced drug candidate sets enriched for migraine-linked pharmacology relative to a global background. Framed in this way, G2DR is best understood as a structured computational engine for narrowing the candidate space for downstream experimental and translational follow-up.

Several features of the results are especially informative. First, transcriptome-prediction resources were only moderately concordant, reinforcing that genetically predicted expression is itself model-dependent and that no single reference architecture should be assumed sufficient. Second, component-level benchmarking showed that different evidence layers optimize different objectives: effect-based ranking was strongest for held-out replication, whereas pathway and hub scores more strongly concentrated curated migraine biology. The integrated score should therefore be interpreted as a balanced downstream-selection score rather than as a universally dominant ranking. Third, contextual comparison with Open Targets suggests that G2DR is complementary to established evidence-integration platforms: it broadens the exploratory target space and recovers additional migraine-reference genes outside the Open Targets migraine list, while remaining less concentrated within established disease-target space.

The drug-level results are most informative when interpreted through the tiered benchmark rather than the global-background enrichment alone. While the framework recovered migraine-linked compounds above random expectation, recovery was strongest for broader literature-linked, off-label, and shared-mechanism therapeutic space, whereas migraine-specific approved therapies were under-represented in the top-ranked set. This suggests that the current implementation is better at expanding genetics-anchored translational hypothesis space than at reproducing the most indication-specific modern pharmacology. Directionality filtering further strengthened this interpretation by showing that only a subset of recovered compounds retained clear mechanistic compatibility with the inferred disease-associated gene direction. Together, these findings argue that the strongest present use case for G2DR is structured target and drug-hypothesis prioritization rather than direct inference of therapeutic efficacy.

The causal-support analyses are important precisely because they limit overinterpretation. Single-instrument Mendelian randomization yielded nominal associations only in broader ranked sets, no gene survived multiple-testing correction, and colocalization identified no gene-locus pairs with strong shared-signal support. These findings do not negate the prioritization signal, but they do indicate that the current framework is operating at the level of genetics-anchored candidate ranking rather than causal effector-gene identification. In practical terms, G2DR should therefore be viewed as a hypothesis-prioritization framework whose outputs require orthogonal validation, not as a causal target-discovery engine.

Across cross-validation folds, exact overlap among top-ranked genes was limited, which is expected in a modest, class-imbalanced cohort in which each fold contains only a small number of cases. More importantly, prioritization performance was stable at the signal level, and drug-level rankings were substantially more stable than gene-level identities. This distinction is central to interpreting the framework correctly: for a prioritization method, reproducibility of ranking behavior is more informative than perfect recurrence of specific gene lists. The results therefore support stability of the prioritization signal even when individual top-ranked genes vary across folds.

The study has several important limitations. The migraine cohort is modest and class-imbalanced, which constrains power and likely contributes to both instability in individual gene identities and diffuse significance across highly layered testing structures. The framework relies on genetically predicted rather than measured context-specific transcription, and TWAS-style signal remains vulnerable to linkage disequilibrium and co-regulation. Drug mapping is further constrained by database coverage, evidence heterogeneity, and imperfect action-direction annotation. These limitations mean that prioritized genes should be interpreted as genetically supported candidates rather than confirmed causal effectors, and prioritized compounds should be interpreted as computational hypotheses rather than translational recommendations.

Equally informative is what the framework does not yet recover well. The under-representation of serotonergic and CGRP-axis therapies suggests that current public drug-target resources and genetically prioritized gene programs more readily capture shared neurobiological and comorbidity-linked pharmacology than highly indication-specific migraine therapeutics. This operating profile helps explain why broader mechanism-linked candidates are enriched even when migraine-specific approved therapies are sparse in the top-ranked set. Future extensions should therefore focus on stronger causal filters, more explicit perturbational and mechanism-of-action resources, and disease-specific pharmacological annotations that may improve both target specificity and directionality-aware drug ranking.

Overall, G2DR provides a useful foundation for genetics-informed prioritization in genotype-first settings. Its modular design allows independent refinement of transcriptome imputation, gene-level ranking, biological contextualization, and drug mapping, and its present value lies in providing a transparent framework for moving from diffuse inherited signal to a more tractable set of experimentally testable hypotheses. The next critical step is not simply to scale the pipeline, but to strengthen causal attribution, benchmark it more directly against alternative genetics-to-drug workflows, and improve translational specificity through external validation and richer pharmacological annotation. Framed in this way, G2DR is a proof-of-concept prioritization framework with clear room for maturation rather than a finished repurposing engine.

\begin{mdframed}[linewidth=1pt,linecolor=black,
innerleftmargin=8pt,innerrightmargin=8pt,
innertopmargin=16pt-8.2pt,innerbottommargin=6pt]
{\fontsize{8.2pt}{10pt}\bfseries Key Points\par}
\begin{adjustwidth}{8pt}{0cm}
\begin{itemize}
\item G2DR is a genotype-first computational prioritization framework for hypothesis generation in settings where matched disease transcriptomics are unavailable; it is intended to narrow the space for downstream experimental and translational follow-up rather than to provide clinically actionable target recommendations directly.
\item The framework integrates genetically predicted gene expression across seven transcriptome-weight resources, multi-method gene-level testing, pathway enrichment, network context, druggability, and multi-source drug--target evidence into a modular and reproducible prioritization pipeline.
\item In held-out internal evaluation, discovery-based gene prioritization generalized to test data (ROC-AUC\,$=0.775$; PR-AUC\,$=0.475$), while the integrated score provided a balanced operating point for downstream target selection rather than maximizing any single benchmark criterion.
\item Drug mapping recovered migraine-linked compounds relative to a global background, but the strongest signal lay in broader mechanism-linked and off-label therapeutic space rather than migraine-specific approved therapies; directionality filtering helped distinguish mechanistically coherent hypotheses from broadly associated candidates.
\item All prioritized genes and compounds remain computational hypotheses requiring independent pharmacological, mechanistic, and clinical validation; in its current form, G2DR is best viewed as a proof-of-concept framework for genetics-anchored prioritization with clear directions for improvement in causal attribution and translational specificity.
\end{itemize}
\end{adjustwidth}
\end{mdframed}

\clearpage

\section{Competing interests}
The authors declare no competing interests.
\section{Author contributions statement}
M.M. conceived the study, designed and implemented the G2DR framework, performed all analyses, and drafted the manuscript. D.B.A. supervised the project, contributed to study design and interpretation, and reviewed and edited the manuscript. All authors approved the final manuscript.
\section{Funding}
This work was supported by the University of Queensland and the Baker Heart and Diabetes Institute. The UK Biobank data were accessed under application ID 50000. The funding sources had no role in study design, data collection, analysis, interpretation, or the decision to submit for publication.
\section{Data and Software Availability} The genotype and phenotype data used in this study were accessed through the UK Biobank under application ID 50000 (\url{https://www.ukbiobank.ac.uk/}) and are subject to UK Biobank access restrictions; researchers may apply for access through the UK Biobank Access Management System. The G2DR framework source code is freely available for non-commercial use at \url{https://github.com/MuhammadMuneeb007/G2DR-A-Genotype-First-Framework-for-Genetics-Informed-Target-Prioritization-and-Drug-Repurposing}. Supporting utilities are available at the following repositories: gene phenotype downloader, \texttt{PhenotypeToGeneDownloaderR} (\url{https://github.com/MuhammadMuneeb007/PhenotypeToGeneDownloaderR}); gene identifier converter, \texttt{GeneMapKit} (\url{https://github.com/MuhammadMuneeb007/GeneMapKit}); drug--disease downloader, \texttt{DownloadDrugsRelatedToDiseases} (\url{https://github.com/MuhammadMuneeb007/DownloadDrugsRelatedToDiseases}). All repositories will be maintained for a minimum of two years following publication. Public annotation resources used by the pipeline include STRING (\url{https://string-db.org}), Open Targets (\url{https://www.opentargets.org}), DGIdb (\url{https://www.dgidb.org}), and ChEMBL (\url{https://www.ebi.ac.uk/chembl}).

\section{Acknowledgments}
Not applicable
\section{Author Biographies} 

\textbf{Muhammad Muneeb} is a PhD candidate at the University of Queensland and Baker Heart and Diabetes Institute, specializing in computational biology.

\textbf{David B. Ascher} is a Professor at the University of Queensland and Baker Heart and Diabetes Institute, leading research in computational structural biology, molecular pharmacology, and translational bioinformatics for drug discovery and precision medicine.
\bibliographystyle{unsrt}
\bibliography{reference}

@article{Martinez2015DrugNet,
  title={DrugNet: network-based drug–disease prioritization},
  author={Martínez, Víctor and et al.},
  journal={Bioinformatics},
  year={2015},
  doi={10.1093/bioinformatics/btv189}
}

@article{Nelson2015,
  title   = {The support of human genetic evidence for approved drug indications},
  author  = {Nelson, Matthew R. and Tipney, Harriet and Painter, Jeffrey L. and Shen, Jessica and Nicoletti, Paolo and Shen, Yufeng and Floratos, Aris and Sham, Pak C. and Li, Michael J. and Wang, Jing and others},
  journal = {Nature Genetics},
  year    = {2015},
  volume  = {47},
  number  = {8},
  pages   = {856--860},
  doi     = {10.1038/ng.3314},
  url     = {https://doi.org/10.1038/ng.3314}
}

@article{araujo2023,
  title = {Multivariate adaptive shrinkage improves cross-population transcriptome prediction and association studies in underrepresented populations},
  author = {Araujo, Daniel S and Nguyen, Chris and Hu, Xiaowei and Mikhaylova, Anna V and Gignoux, Christopher and Ardlie, Kristin and Taylor, Kent D and Durda, Peter and Liu, Yongmei and Papanicolaou, George and others},
  journal = {HGG advances},
  volume = {4},
  number = {4},
  pages = {100250},
  year = {2023},
  doi = {10.1016/j.xhgg.2023.100250}
}

@article{zhou2020unified,
  title = {A unified framework for joint-tissue transcriptome-wide association and Mendelian randomization analysis},
  author = {Zhou, Dan and Jiang, Yi and Zhong, Xiaohong and Cox, Nancy J and Liu, Chunyu and Gamazon, Eric R},
  journal = {Nature genetics},
  volume = {52},
  number = {11},
  pages = {1234--1242},
  year = {2020},
  doi = {10.1038/s41588-020-0706-2}
}

@article{zhang2019integrative,
  title = {Integrative transcriptome imputation reveals tissue-specific and shared biological mechanisms mediating susceptibility to complex traits},
  author = {Zhang, Wen and Voloudakis, Georgios and Rajagopal, Veera M and Readhead, Ben and Dudley, Joel T and Schadt, Eric E and Bj{\"o}rkegren, Johan L M and Kim, Yungil and Fullard, John F and Hoffman, Gabriel E and others},
  journal = {Nature communications},
  volume = {10},
  number = {1},
  pages = {3834},
  year = {2019},
  doi = {10.1038/s41467-019-11874-7}
}

@article{gusev2016integrative,
  title = {Integrative approaches for large-scale transcriptome-wide association studies},
  author = {Gusev, Alexander and Ko, Arthur and Shi, Huwenbo and Bhatia, Gaurav and Chung, Wendy and Penninx, Brenda WJH and Jansen, Rick and De Geus, Eco JC and Boomsma, Dorret I and Wright, Fred A and others},
  journal = {Nature genetics},
  volume = {48},
  number = {3},
  pages = {245--252},
  year = {2016},
  doi = {10.1038/ng.3506}
}

@article{nagpal2019tigar,
  title = {TIGAR: an improved Bayesian tool for transcriptomic data imputation enhances gene mapping of complex traits},
  author = {Nagpal, Sini and Meng, Xing and Epstein, Michael P and Tsoi, Lam C and Patrick, Matthew and Gibson, Greg and De Jager, Philip L and Bennett, David A and Wingo, Thomas S and Wingo, Aliza P and others},
  journal = {The American Journal of Human Genetics},
  volume = {105},
  number = {2},
  pages = {258--266},
  year = {2019},
  doi = {10.1016/j.ajhg.2019.05.018}
}

@article{ritchie2015limma,
  title={limma powers differential expression analyses for RNA-sequencing and microarray studies},
  author={Ritchie, Matthew E and Phipson, Belinda and Wu, Di and others},
  journal={Nucleic Acids Research},
  volume={43},
  number={7},
  pages={e47--e47},
  year={2015},
  publisher={Oxford University Press},
	doi = {10.1093/nar/gkv007},
}

@article{Bigal2009Comorbidity,
  title={Migraine and cardiovascular disease: a population-based study},
  author={Bigal, Marcelo E. and Lipton, Richard B.},
  journal={Neurology},
  volume={72},
  number={21},
  pages={1864--1871},
  year={2009},
  doi={10.1212/WNL.0b013e3181a71220}
}

@article{Gazerani2015GutBrain,
  title={Migraine and gastrointestinal disorders: a systematic review},
  author={Gazerani, Parisa},
  journal={The Journal of Headache and Pain},
  volume={16},
  pages={39},
  year={2015},
  doi={10.1186/s10194-015-0523-7}
}

@article{Katsarava2018Migraine,
  title={Migraine and comorbidities},
  author={Katsarava, Zaza and Buse, Dawn C. and Manack, Adam N. and Lipton, Richard B.},
  journal={The Journal of Headache and Pain},
  volume={19},
  number={1},
  pages={126},
  year={2018},
  doi={10.1186/s10194-018-0934-8}
}

@article{Buse2013Comorbidity,
  title={Migraine-related disability, impact, and health-related quality of life},
  author={Buse, Dawn C and Manack, Adam and Serrano, Daniel and Turkel, Connie and Lipton, Richard B},
  journal={Neurology},
  volume={80},
  number={24},
  pages={2194--2203},
  year={2013},
  publisher={American Academy of Neurology}
}

@article{Purcell2007PLINK,
  title={PLINK: a tool set for whole-genome association and population-based linkage analyses},
  author={Purcell, Shaun and Neale, Benjamin and Todd-Brown, Kathryn and Thomas, Lori and Ferreira, Manuel AR and Bender, David and Maller, Julian and Sklar, Pamela and de Bakker, Paul IW and Daly, Mark J and Sham, Pak C},
  journal={American Journal of Human Genetics},
  volume={81},
  number={3},
  pages={559--575},
  year={2007},
  publisher={Elsevier},
	doi = {10.1086/519795},
}

@article{Anderson2010QC,
  title={Data quality control in genetic case-control association studies},
  author={Anderson, Carl A and Pettersson, Fredrik H and Clarke, Gerald M and Cardon, Lon R and Morris, Andrew P and Zondervan, Krina T},
  journal={Nature Protocols},
  volume={5},
  number={9},
  pages={1564--1573},
  year={2010},
  publisher={Nature Publishing Group},
	doi = {10.1038/nprot.2010.116},
}

@article{Wishart2018,
  title={DrugBank 5.0: a major update to the DrugBank database},
  author={Wishart, David S and Feunang, Yannick D and Guo, An C and others},
  journal={Nucleic Acids Research},
  volume={46},
  number={D1},
  pages={D1074--D1082},
  year={2018},
  publisher={Oxford University Press}
}

@article{Kirthi2013,
  author  = {Kirthi, V. and Derry, S. and Moore, R. A.},
  title   = {Aspirin with or without an antiemetic for acute migraine headaches in adults},
  journal = {Cochrane Database of Systematic Reviews},
  year    = {2013},
  doi     = {10.1002/14651858.CD008041.pub3}
}

@article{Pomeroy2017,
  author  = {Pomeroy, J. L. and Marmura, M. J. and Nahas, S. J. and Viscusi, E. R.},
  title   = {Ketamine Infusions for Treatment Refractory Headache},
  journal = {Headache},
  volume  = {57},
  number  = {2},
  pages   = {276--282},
  year    = {2017},
  doi     = {10.1111/head.13013}
}

@article{Derry2013,
  title = {Diclofenac with or without an antiemetic for acute migraine headaches in adults},
  volume = {2019},
  ISSN = {1465-1858},
  url = {http://dx.doi.org/10.1002/14651858.CD008783.pub3},
  DOI = {10.1002/14651858.cd008783.pub3},
  number = {5},
  journal = {Cochrane Database of Systematic Reviews},
  publisher = {Wiley},
  author = {Derry,  Sheena and Rabbie,  Roy and Moore,  R Andrew},
  year = {2013},
  month = apr 
}

@article{Marmura2015,
  author  = {Marmura, M. J. and Goldberg, S. W.},
  title   = {Celecoxib for acute migraine: a systematic review},
  journal = {Headache},
  volume  = {55},
  number  = {3},
  pages   = {387--393},
  year    = {2015},
  doi     = {10.1111/head.12518}
}

@article{Singh2008,
  author  = {Singh, A. and Alter, H. J. and Zaia, B.},
  title   = {Does the addition of dexamethasone to standard therapy for acute migraine headache decrease the incidence of recurrent headache for patients treated in the emergency department? A meta-analysis and systematic review of the literature},
  journal = {Academic Emergency Medicine},
  volume  = {15},
  number  = {12},
  pages   = {1223--1233},
  year    = {2008},
  doi     = {10.1111/j.1553-2712.2008.00283.x}
}

@article{Colman2008,
  author  = {Colman, I. and Friedman, B. W. and Brown, M. D. and Innes, G. D. and Grafstein, E. and Roberts, T. E. and Rowe, B. H.},
  title   = {Parenteral dexamethasone for acute severe migraine headache: meta-analysis of randomised controlled trials for preventing recurrence},
  journal = {BMJ},
  volume  = {336},
  number  = {7657},
  pages   = {1359--1361},
  year    = {2008},
  doi     = {10.1136/bmj.39566.806725.BE}
}

@article{Crivellaro2021,
  title   = {Specific activation of GluN1-N2B NMDA receptors underlies facilitation of cortical spreading depression in a genetic mouse model of migraine with reduced astrocytic glutamate clearance},
  author  = {Crivellaro, Giovanna and others},
  journal = {Neurobiology of Disease},
  year    = {2021},
  volume  = {156},
  pages   = {105419},
  doi     = {10.1016/j.nbd.2021.105419},
  pmid    = {34111520}
}

@article{Pushpakom2019,
  title   = {Drug repurposing: progress, challenges and recommendations},
  author  = {Pushpakom, Sudeep and Iorio, Francesco and Eyers, Patrick A. and Escott, Kieron J. and Hopper, Shirley and Wells, Andrew and Doig, Andrew and Guilliams, Tim and Latimer, Joanna and McNamee, Catherine and Norris, Alison and Sanseau, Philippe and Cavalla, Daniela and Pirmohamed, Munir},
  journal = {Nature Reviews Drug Discovery},
  year    = {2019},
  volume  = {18},
  number  = {1},
  pages   = {41--58},
  doi     = {10.1038/s41573-019-0040-6},
  url     = {https://doi.org/10.1038/s41573-019-0040-6}
}

@article{Lamb2006,
  title   = {The Connectivity Map: using gene-expression signatures to connect small molecules, genes, and disease},
  author  = {Lamb, Justin and Crawford, Emily D. and Peck, David and Modell, Joshua W. and Blat, Irene C. and Wrobel, Matthew J. and Lerner, Jim and Brunet, Jean-Philippe and Subramanian, Aravind and Ross, Kenneth N. and Reich, Michael and Hieronymus, Haley and Wei, Guo and Armstrong, Scott A. and Haggarty, Stephen J. and Clemons, Paul A. and Wei, Ru and Carr, Steven A. and Lander, Eric S. and Golub, Todd R.},
  journal = {Science},
  year    = {2006},
  volume  = {313},
  number  = {5795},
  pages   = {1929--1935},
  doi     = {10.1126/science.1132939},
  url     = {https://doi.org/10.1126/science.1132939}
}

@article{Gamazon2015,
  title   = {A gene-based association method for mapping traits using reference transcriptome data},
  author  = {Gamazon, Eric R. and Wheeler, Heather E. and Shah, Kaan P. and Mozaffari, Somayeh V. and Aquino-Michaels, Karolina and Carroll, Rory J. and Eyler, Anne E. and Denny, Joshua C. and Nicolae, Dan L. and Cox, Nancy J. and Im, Hong-Hee},
  journal = {Nature Genetics},
  year    = {2015},
  volume  = {47},
  number  = {9},
  pages   = {1091--1098},
  doi     = {10.1038/ng.3367},
  url     = {https://doi.org/10.1038/ng.3367}
}

@article{Wainberg2019,
  title   = {Opportunities and challenges for transcriptome-wide association studies},
  author  = {Wainberg, Michael and Sinnott-Armstrong, Nasa and Mancuso, Nicolo and Barbeira, Alvaro N. and Knowles, David A. and Golan, David and Ermel, Rachel and Ruusalepp, Arno and Quertermous, Thomas and Hao, Ke and Bj{\"o}rkegren, Johan L.M. and Im, Hong-Hee and Pasaniuc, Bogdan and Rivas, Manuel A.},
  journal = {Nature Genetics},
  year    = {2019},
  volume  = {51},
  number  = {4},
  pages   = {592--599},
  doi     = {10.1038/s41588-019-0385-x},
  url     = {https://doi.org/10.1038/s41588-019-0385-x}
}

@article{Ashburn2004,
  title = {Drug repositioning: identifying and developing new uses for existing drugs},
  volume = {3},
  ISSN = {1474-1784},
  url = {http://dx.doi.org/10.1038/nrd1468},
  DOI = {10.1038/nrd1468},
  number = {8},
  journal = {Nature Reviews Drug Discovery},
  publisher = {Springer Science and Business Media LLC},
  author = {Ashburn,  Ted T. and Thor,  Karl B.},
  year = {2004},
  month = aug,
  pages = {673–683}
}

@article{Nosengo2016,
  title={Can you teach old drugs new tricks?},
  author={Nosengo, Nicola},
  journal={Nature},
  volume={534},
  pages={314--316},
  year={2016},
  publisher={Nature Publishing Group},
	doi = {10.1038/534314a},
}

@article{Koleti2018,
  title={Data Portal for the Library of Integrated Network-Based Cellular Signatures (LINCS) program},
  author={Koleti, Amar and Terryn, R and Stathias, V and others},
  journal={Nucleic Acids Research},
  volume={46},
  number={D1},
  pages={D558--D566},
  year={2018},
  publisher={Oxford University Press}
}

@article{Hopkins2008,
  title={Network pharmacology: the next paradigm in drug discovery},
  author={Hopkins, Andrew L},
  journal={Nature Chemical Biology},
  volume={4},
  number={11},
  pages={682--690},
  year={2008},
  publisher={Nature Publishing Group},
	doi = {10.1038/nchembio.118},
}

@article{Zitnik2018,
  title={Modeling polypharmacy side effects with graph convolutional networks},
  author={Zitnik, Marinka and Agrawal, Monica and Leskovec, Jure},
  journal={Bioinformatics},
  volume={34},
  number={13},
  pages={i457--i466},
  year={2018},
  publisher={Oxford University Press},
	doi = {10.1101/258814},
}

@article{Ochoa2021,
  title={The Open Targets Platform: supporting systematic drug--target identification and prioritisation},
  author={Ochoa, David and Hercules, Andrew and Carmona, Miguel and others},
  journal={Nucleic Acids Research},
  volume={49},
  number={D1},
  pages={D1302--D1310},
  year={2021},
  publisher={Oxford University Press},
	doi = {10.1093/nar/gkaa1027},
}

@article{Freshour2021,
  title={Integration of the Drug--Gene Interaction Database (DGIdb 4.0)},
  author={Freshour, Summer L and Kiwala, Savannah and Cotto, Kelsy C and others},
  journal={Nucleic Acids Research},
  volume={49},
  number={D1},
  pages={D1144--D1151},
  year={2021},
  publisher={Oxford University Press}
}

@article{Subramanian2017,
  title        = {A Next Generation Connectivity Map: {L}1000 Platform and the First 1,000,000 Profiles},
  author       = {Subramanian, Aravind and Narayan, Rohith and Corsello, Steven M. and Peck, David D. and Natoli, Theodore E. and Lu, Xiaodong and Gould, Joshua and Davis, John F. and Tubelli, Alberto A. and Asiedu, Joseph K. and Lahr, David L. and Hirschman, Joseph E. and Liu, Zihan and Donahue, Michael and Julian, Brian and Khan, Mehrtash and Wadden, David and Smith, Ian C. and Lam, Daniel and Liberzon, Arthur and Toder, Craig and Bagul, Megan and Orzechowski, Marcin and Enache, Oana M. and Piccioni, Francesco and Johnson, Scott A. and Lyons, Nicholas J. and Berger, Adam H. and Shamji, Alykhan F. and Brooks, Aaron N. and Vrcic, Anita and Flynn, Colleen and Rosains, Jenna and Takeda, Daisuke Y. and Hu, Rong and Davison, Daniel and Lamb, Justin and Ardlie, Kristin and Hogstrom, Larson and Greenside, Peyton and Gray, Nathanael S. and Clemons, Paul A. and Silver, Seth and Wu, Xiaoyun and Zhao, W. Nicholas and Read-Button, William and Wu, Xiao and Haggarty, Stephen J. and Ronco, Luigi V. and Boehm, Jesse S. and Schreiber, Stuart L. and Doench, John G. and Bittker, Joshua A. and Root, David E. and Wong, Bang and Golub, Todd R.},
  journal      = {Cell},
  year         = {2017},
  volume       = {171},
  number       = {6},
  pages        = {1437--1452.e17},
  doi          = {10.1016/j.cell.2017.10.049}
}

@article{Himmelstein2017,
  title        = {Systematic integration of biomedical knowledge prioritizes drugs for repurposing},
  author       = {Himmelstein, Daniel S. and Lizee, Antoine and Hessler, Christine and Brueggeman, Leo and Chen, Sabrina L. and Hadley, Dexter and Green, Ari and Khankhanian, Pouya and Baranzini, Sergio E.},
  journal      = {eLife},
  year         = {2017},
  volume       = {6},
  pages        = {e26726},
  doi          = {10.7554/eLife.26726}
}

@article{Barabasi2011,
  title        = {Network medicine: a network-based approach to human disease},
  author       = {Barab{\'a}si, Albert-L{\'a}szl{\'o} and Gulbahce, Natali and Loscalzo, Joseph},
  journal      = {Nature Reviews Genetics},
  year         = {2011},
  volume       = {12},
  number       = {1},
  pages        = {56--68},
  doi          = {10.1038/nrg2918}
}

@article{Menche2015,
  title        = {Uncovering disease--disease relationships through the incomplete interactome},
  author       = {Menche, J{\"o}rg and Sharma, Amitabh and Kitsak, Maksim and Ghiassian, Susan D. and Vidal, Marc and Loscalzo, Joseph and Barab{\'a}si, Albert-L{\'a}szl{\'o}},
  journal      = {Science},
  year         = {2015},
  volume       = {347},
  number       = {6224},
  pages        = {1257601},
  doi          = {10.1126/science.1257601}
}

@article{Sudlow2015,
  title        = {{UK} {B}iobank: an open access resource for identifying the causes of a wide range of complex diseases of middle and old age},
  author       = {Sudlow, Cathie and Gallacher, John and Allen, Naomi and Beral, Valerie and Burton, Paul and Danesh, John and Downey, Paul and Elliott, Paul and Green, Jane and Landray, Martin and Liu, Bette and Matthews, Paul and Ong, Gideon and Pell, Jill and Silman, Alan and Young, Alan and Sprosen, Tim and Peakman, Tim and Collins, Rory},
  journal      = {PLOS Medicine},
  year         = {2015},
  volume       = {12},
  number       = {3},
  pages        = {e1001779},
  doi          = {10.1371/journal.pmed.1001779}
}

@article{Bycroft2018,
  title        = {The {UK} {B}iobank resource with deep phenotyping and genomic data},
  author       = {Bycroft, Clare and Freeman, Colin and Petkova, Desislava and Band, Gavin and Elliott, Lloyd T. and Sharp, Kevin and Motyer, Allan and Vukcevic, Damjan and Delaneau, Olivier and O'Connell, Jared and Cortes, Adrian and Welsh, Samantha and McVean, Gil and Leslie, Stephen and Donnelly, Peter and Marchini, Jonathan},
  journal      = {Nature},
  year         = {2018},
  volume       = {562},
  number       = {7726},
  pages        = {203--209},
  doi          = {10.1038/s41586-018-0579-z}
}

@article{King2019,
  title        = {Are drug targets with genetic support twice as likely to be approved? Revised estimates of the impact of genetic support for drug mechanisms on the probability of drug approval},
  author       = {King, Emily A. and Davis, Jimmy W. and Degner, Jacob F.},
  journal      = {PLOS Genetics},
  year         = {2019},
  volume       = {15},
  number       = {12},
  pages        = {e1008489},
  doi          = {10.1371/journal.pgen.1008489}
}

@article{CTIMP,
  title = {A statistical framework for cross-tissue transcriptome-wide association analysis},
  volume = {51},
  ISSN = {1546-1718},
  url = {http://dx.doi.org/10.1038/s41588-019-0345-7},
  DOI = {10.1038/s41588-019-0345-7},
  number = {3},
  journal = {Nature Genetics},
  publisher = {Springer Science and Business Media LLC},
  author = {Hu,  Yiming and Li,  Mo and Lu,  Qiongshi and Weng,  Haoyi and Wang,  Jiawei and Zekavat,  Seyedeh M. and Yu,  Zhaolong and Li,  Boyang and Gu,  Jianlei and Muchnik,  Sydney and Shi,  Yu and Kunkle,  Brian W. and Mukherjee,  Shubhabrata and Natarajan,  Pradeep and Naj,  Adam and Kuzma,  Amanda and Zhao,  Yi and Crane,  Paul K. and Lu,  Hui and Zhao,  Hongyu},
  year = {2019},
  month = feb,
  pages = {568–576}
}

@article{Barbeira2018,
  title = {Exploring the phenotypic consequences of tissue specific gene expression variation inferred from GWAS summary statistics},
  volume = {9},
  ISSN = {2041-1723},
  url = {http://dx.doi.org/10.1038/s41467-018-03621-1},
  DOI = {10.1038/s41467-018-03621-1},
  number = {1},
  journal = {Nature Communications},
  publisher = {Springer Science and Business Media LLC},
  author = {Barbeira,  Alvaro N. and Dickinson,  Scott P. and Bonazzola,  Rodrigo and Zheng,  Jiamao and Wheeler,  Heather E. and Torres,  Jason M. and Torstenson,  Eric S. and Shah,  Kaanan P. and Garcia,  Tzintzuni and Edwards,  Todd L. and Stahl,  Eli A. and Huckins,  Laura M. and Aguet,  Fran\c{c}ois and Ardlie,  Kristin G. and Cummings,  Beryl B. and Gelfand,  Ellen T. and Getz,  Gad and Hadley,  Kane and Handsaker,  Robert E. and Huang,  Katherine H. and Kashin,  Seva and Karczewski,  Konrad J. and Lek,  Monkol and Li,  Xiao and MacArthur,  Daniel G. and Nedzel,  Jared L. and Nguyen,  Duyen T. and Noble,  Michael S. and Segrè,  Ayellet V. and Trowbridge,  Casandra A. and Tukiainen,  Taru and Abell,  Nathan S. and Balliu,  Brunilda and Barshir,  Ruth and Basha,  Omer and Battle,  Alexis and Bogu,  Gireesh K. and Brown,  Andrew and Brown,  Christopher D. and Castel,  Stephane E. and Chen,  Lin S. and Chiang,  Colby and Conrad,  Donald F. and Damani,  Farhan N. and Davis,  Joe R. and Delaneau,  Olivier and Dermitzakis,  Emmanouil T. and Engelhardt,  Barbara E. and Eskin,  Eleazar and Ferreira,  Pedro G. and Frésard,  Laure and Gamazon,  Eric R. and Garrido-Martín,  Diego and Gewirtz,  Ariel D. H. and Gliner,  Genna and Gloudemans,  Michael J. and Guigo,  Roderic and Hall,  Ira M. and Han,  Buhm and He,  Yuan and Hormozdiari,  Farhad and Howald,  Cedric and Jo,  Brian and Kang,  Eun Yong and Kim,  Yungil and Kim-Hellmuth,  Sarah and Lappalainen,  Tuuli and Li,  Gen and Li,  Xin and Liu,  Boxiang and Mangul,  Serghei and McCarthy,  Mark I. and McDowell,  Ian C. and Mohammadi,  Pejman and Monlong,  Jean and Montgomery,  Stephen B. and Muñoz-Aguirre,  Manuel and Ndungu,  Anne W. and Nobel,  Andrew B. and Oliva,  Meritxell and Ongen,  Halit and Palowitch,  John J. and Panousis,  Nikolaos and Papasaikas,  Panagiotis and Park,  YoSon and Parsana,  Princy and Payne,  Anthony J. and Peterson,  Christine B. and Quan,  Jie and Reverter,  Ferran and Sabatti,  Chiara and Saha,  Ashis and Sammeth,  Michael and Scott,  Alexandra J. and Shabalin,  Andrey A. and Sodaei,  Reza and Stephens,  Matthew and Stranger,  Barbara E. and Strober,  Benjamin J. and Sul,  Jae Hoon and Tsang,  Emily K. and Urbut,  Sarah and van de Bunt,  Martijn and Wang,  Gao and Wen,  Xiaoquan and Wright,  Fred A. and Xi,  Hualin S. and Yeger-Lotem,  Esti and Zappala,  Zachary and Zaugg,  Judith B. and Zhou,  Yi-Hui and Akey,  Joshua M. and Bates,  Daniel and Chan,  Joanne and Chen,  Lin S. and Claussnitzer,  Melina and Demanelis,  Kathryn and Diegel,  Morgan and Doherty,  Jennifer A. and Feinberg,  Andrew P. and Fernando,  Marian S. and Halow,  Jessica and Hansen,  Kasper D. and Haugen,  Eric and Hickey,  Peter F. and Hou,  Lei and Jasmine,  Farzana and Jian,  Ruiqi and Jiang,  Lihua and Johnson,  Audra and Kaul,  Rajinder and Kellis,  Manolis and Kibriya,  Muhammad G. and Lee,  Kristen and Li,  Jin Billy and Li,  Qin and Li,  Xiao and Lin,  Jessica and Lin,  Shin and Linder,  Sandra and Linke,  Caroline and Liu,  Yaping and Maurano,  Matthew T. and Molinie,  Benoit and Montgomery,  Stephen B. and Nelson,  Jemma and Neri,  Fidencio J. and Oliva,  Meritxell and Park,  Yongjin and Pierce,  Brandon L. and Rinaldi,  Nicola J. and Rizzardi,  Lindsay F. and Sandstrom,  Richard and Skol,  Andrew and Smith,  Kevin S. and Snyder,  Michael P. and Stamatoyannopoulos,  John and Stranger,  Barbara E. and Tang,  Hua and Tsang,  Emily K. and Wang,  Li and Wang,  Meng and Van Wittenberghe,  Nicholas and Wu,  Fan and Zhang,  Rui and Nierras,  Concepcion R. and Branton,  Philip A. and Carithers,  Latarsha J. and Guan,  Ping and Moore,  Helen M. and Rao,  Abhi and Vaught,  Jimmie B. and Gould,  Sarah E. and Lockart,  Nicole C. and Martin,  Casey and Struewing,  Jeffery P. and Volpi,  Simona and Addington,  Anjene M. and Koester,  Susan E. and Little,  A. Roger and Brigham,  Lori E. and Hasz,  Richard and Hunter,  Marcus and Johns,  Christopher and Johnson,  Mark and Kopen,  Gene and Leinweber,  William F. and Lonsdale,  John T. and McDonald,  Alisa and Mestichelli,  Bernadette and Myer,  Kevin and Roe,  Brian and Salvatore,  Michael and Shad,  Saboor and Thomas,  Jeffrey A. and Walters,  Gary and Washington,  Michael and Wheeler,  Joseph and Bridge,  Jason and Foster,  Barbara A. and Gillard,  Bryan M. and Karasik,  Ellen and Kumar,  Rachna and Miklos,  Mark and Moser,  Michael T. and Jewell,  Scott D. and Montroy,  Robert G. and Rohrer,  Daniel C. and Valley,  Dana R. and Davis,  David A. and Mash,  Deborah C. and Undale,  Anita H. and Smith,  Anna M. and Tabor,  David E. and Roche,  Nancy V. and McLean,  Jeffrey A. and Vatanian,  Negin and Robinson,  Karna L. and Sobin,  Leslie and Barcus,  Mary E. and Valentino,  Kimberly M. and Qi,  Liqun and Hunter,  Steven and Hariharan,  Pushpa and Singh,  Shilpi and Um,  Ki Sung and Matose,  Takunda and Tomaszewski,  Maria M. and Barker,  Laura K. and Mosavel,  Maghboeba and Siminoff,  Laura A. and Traino,  Heather M. and Flicek,  Paul and Juettemann,  Thomas and Ruffier,  Magali and Sheppard,  Dan and Taylor,  Kieron and Trevanion,  Stephen J. and Zerbino,  Daniel R. and Craft,  Brian and Goldman,  Mary and Haeussler,  Maximilian and Kent,  W. James and Lee,  Christopher M. and Paten,  Benedict and Rosenbloom,  Kate R. and Vivian,  John and Zhu,  Jingchun and Nicolae,  Dan L. and Cox,  Nancy J. and Im,  Hae Kyung},
  year = {2018},
  month = may 
}

@article{Gaulton2011,
  title = {ChEMBL: a large-scale bioactivity database for drug discovery},
  volume = {40},
  ISSN = {1362-4962},
  url = {http://dx.doi.org/10.1093/nar/gkr777},
  DOI = {10.1093/nar/gkr777},
  number = {D1},
  journal = {Nucleic Acids Research},
  publisher = {Oxford University Press (OUP)},
  author = {Gaulton,  A. and Bellis,  L. J. and Bento,  A. P. and Chambers,  J. and Davies,  M. and Hersey,  A. and Light,  Y. and McGlinchey,  S. and Michalovich,  D. and Al-Lazikani,  B. and Overington,  J. P.},
  year = {2011},
  month = sep,
  pages = {D1100–D1107}
}

@article{Gaulton2016,
  title = {The ChEMBL database in 2017},
  volume = {45},
  ISSN = {1362-4962},
  url = {http://dx.doi.org/10.1093/nar/gkw1074},
  DOI = {10.1093/nar/gkw1074},
  number = {D1},
  journal = {Nucleic Acids Research},
  publisher = {Oxford University Press (OUP)},
  author = {Gaulton,  Anna and Hersey,  Anne and Nowotka,  Michał and Bento,  A. Patrícia and Chambers,  Jon and Mendez,  David and Mutowo,  Prudence and Atkinson,  Francis and Bellis,  Louisa J. and Cibrián-Uhalte,  Elena and Davies,  Mark and Dedman,  Nathan and Karlsson,  Anneli and Magariños,  María Paula and Overington,  John P. and Papadatos,  George and Smit,  Ines and Leach,  Andrew R.},
  year = {2016},
  month = nov,
  pages = {D945–D954}
}

@article{Schwenk2022,
  title = {Lidocaine infusions for refractory chronic migraine: a retrospective analysis},
  volume = {47},
  ISSN = {1532-8651},
  url = {http://dx.doi.org/10.1136/rapm-2021-103180},
  DOI = {10.1136/rapm-2021-103180},
  number = {7},
  journal = {Regional Anesthesia and; Pain Medicine},
  publisher = {BMJ},
  author = {Schwenk,  Eric S and Walter,  Aaron and Torjman,  Marc C and Mukhtar,  Sarah and Patel,  Harsh T and Nardone,  Bryan and Sun,  George and Thota,  Bhavana and Lauritsen,  Clinton G and Silberstein,  Stephen D},
  year = {2022},
  month = may,
  pages = {408–413}
}

@article{Hautakangas2022,
  title = {Genome-wide analysis of 102, 084 migraine cases identifies 123 risk loci and subtype-specific risk alleles},
  volume = {54},
  ISSN = {1546-1718},
  url = {http://dx.doi.org/10.1038/s41588-021-00990-0},
  DOI = {10.1038/s41588-021-00990-0},
  number = {2},
  journal = {Nature Genetics},
  publisher = {Springer Science and Business Media LLC},
  author = {Hautakangas,  Heidi and Winsvold,  Bendik S. and Ruotsalainen,  Sanni E. and Bjornsdottir,  Gyda and Harder,  Aster V. E. and Kogelman,  Lisette J. A. and Thomas,  Laurent F. and Noordam,  Raymond and Benner,  Christian and Gormley,  Padhraig and Artto,  Ville and Banasik,  Karina and Bjornsdottir,  Anna and Boomsma,  Dorret I. and Brumpton,  Ben M. and Burgdorf,  Kristoffer Sølvsten and Buring,  Julie E. and Chalmer,  Mona Ameri and de Boer,  Irene and Dichgans,  Martin and Erikstrup,  Christian and F\"{a}rkkil\"{a},  Markus and Garbrielsen,  Maiken Elvestad and Ghanbari,  Mohsen and Hagen,  Knut and H\"{a}pp\"{o}l\"{a},  Paavo and Hottenga,  Jouke-Jan and Hrafnsdottir,  Maria G. and Hveem,  Kristian and Johnsen,  Marianne Bakke and K\"{a}h\"{o}nen,  Mika and Kristoffersen,  Espen S. and Kurth,  Tobias and Lehtim\"{a}ki,  Terho and Lighart,  Lannie and Magnusson,  Sigurdur H. and Malik,  Rainer and Pedersen,  Ole Birger and Pelzer,  Nadine and Penninx,  Brenda W. J. H. and Ran,  Caroline and Ridker,  Paul M. and Rosendaal,  Frits R. and Sigurdardottir,  Gudrun R. and Skogholt,  Anne Heidi and Sveinsson,  Olafur A. and Thorgeirsson,  Thorgeir E. and Ullum,  Henrik and Vijfhuizen,  Lisanne S. and Widén,  Elisabeth and van Dijk,  Ko Willems and de Boer,  Irene and van den Maagdenberg,  Arn M. J. M. and Aromaa,  Arpo and Belin,  Andrea Carmine and Freilinger,  Tobias and Ikram,  M. Arfan and J\"{a}rvelin,  Marjo-Riitta and Raitakari,  Olli T. and Terwindt,  Gisela M. and Kallela,  Mikko and Wessman,  Maija and Olesen,  Jes and Chasman,  Daniel I. and Nyholt,  Dale R. and Stefánsson,  Hreinn and Stefansson,  Kari and van den Maagdenberg,  Arn M. J. M. and Hansen,  Thomas Folkmann and Ripatti,  Samuli and Zwart,  John-Anker and Palotie,  Aarno and Pirinen,  Matti},
  year = {2022},
  month = feb,
  pages = {152–160}
}

@article{Gormley2016,
  title = {Meta-analysis of 375, 000 individuals identifies 38 susceptibility loci for migraine},
  volume = {48},
  ISSN = {1546-1718},
  url = {http://dx.doi.org/10.1038/ng.3598},
  DOI = {10.1038/ng.3598},
  number = {8},
  journal = {Nature Genetics},
  publisher = {Springer Science and Business Media LLC},
  author = {Gormley,  Padhraig and Anttila,  Verneri and Winsvold,  Bendik S and Palta,  Priit and Esko,  Tonu and Pers,  Tune H and Farh,  Kai-How and Cuenca-Leon,  Ester and Muona,  Mikko and Furlotte,  Nicholas A and Kurth,  Tobias and Ingason,  Andres and McMahon,  George and Ligthart,  Lannie and Terwindt,  Gisela M and Kallela,  Mikko and Freilinger,  Tobias M and Ran,  Caroline and Gordon,  Scott G and Stam,  Anine H and Steinberg,  Stacy and Borck,  Guntram and Koiranen,  Markku and Quaye,  Lydia and Adams,  Hieab H H and Lehtim\"{a}ki,  Terho and Sarin,  Antti-Pekka and Wedenoja,  Juho and Hinds,  David A and Buring,  Julie E and Sch\"{u}rks,  Markus and Ridker,  Paul M and Hrafnsdottir,  Maria Gudlaug and Stefansson,  Hreinn and Ring,  Susan M and Hottenga,  Jouke-Jan and Penninx,  Brenda W J H and F\"{a}rkkil\"{a},  Markus and Artto,  Ville and Kaunisto,  Mari and Veps\"{a}l\"{a}inen,  Salli and Malik,  Rainer and Heath,  Andrew C and Madden,  Pamela A F and Martin,  Nicholas G and Montgomery,  Grant W and Kurki,  Mitja I and Kals,  Mart and M\"{a}gi,  Reedik and P\"{a}rn,  Kalle and H\"{a}m\"{a}l\"{a}inen,  Eija and Huang,  Hailiang and Byrnes,  Andrea E and Franke,  Lude and Huang,  Jie and Stergiakouli,  Evie and Lee,  Phil H and Sandor,  Cynthia and Webber,  Caleb and Cader,  Zameel and Muller-Myhsok,  Bertram and Schreiber,  Stefan and Meitinger,  Thomas and Eriksson,  Johan G and Salomaa,  Veikko and Heikkil\"{a},  Kauko and Loehrer,  Elizabeth and Uitterlinden,  Andre G and Hofman,  Albert and van Duijn,  Cornelia M and Cherkas,  Lynn and Pedersen,  Linda M and Stubhaug,  Audun and Nielsen,  Christopher S and M\"{a}nnikk\"{o},  Minna and Mihailov,  Evelin and Milani,  Lili and G\"{o}bel,  Hartmut and Esserlind,  Ann-Louise and Christensen,  Anne Francke and Hansen,  Thomas Folkmann and Werge,  Thomas and Kaprio,  Jaakko and Aromaa,  Arpo J and Raitakari,  Olli and Ikram,  M Arfan and Spector,  Tim and J\"{a}rvelin,  Marjo-Riitta and Metspalu,  Andres and Kubisch,  Christian and Strachan,  David P and Ferrari,  Michel D and Belin,  Andrea C and Dichgans,  Martin and Wessman,  Maija and van den Maagdenberg,  Arn M J M and Zwart,  John-Anker and Boomsma,  Dorret I and Smith,  George Davey and Stefansson,  Kari and Eriksson,  Nicholas and Daly,  Mark J and Neale,  Benjamin M and Olesen,  Jes and Chasman,  Daniel I and Nyholt,  Dale R and Palotie,  Aarno},
  year = {2016},
  month = jun,
  pages = {856–866}
}

@article{Gasparini2016,
  title = {Genetic insights into migraine and glutamate: a protagonist driving the headache},
  volume = {367},
  ISSN = {0022-510X},
  url = {http://dx.doi.org/10.1016/j.jns.2016.06.016},
  DOI = {10.1016/j.jns.2016.06.016},
  journal = {Journal of the Neurological Sciences},
  publisher = {Elsevier BV},
  author = {Gasparini,  Claudia F. and Smith,  Robert A. and Griffiths,  Lyn R.},
  year = {2016},
  month = aug,
  pages = {258–268}
}

@article{Gross2019,
  title = {The metabolic face of migraine — from pathophysiology to treatment},
  volume = {15},
  ISSN = {1759-4766},
  url = {http://dx.doi.org/10.1038/s41582-019-0255-4},
  DOI = {10.1038/s41582-019-0255-4},
  number = {11},
  journal = {Nature Reviews Neurology},
  publisher = {Springer Science and Business Media LLC},
  author = {Gross,  Elena C. and Lisicki,  Marco and Fischer,  Dirk and Sándor,  Peter S. and Schoenen,  Jean},
  year = {2019},
  month = oct,
  pages = {627–643}
}

@article{Curto2016,
  title = {Altered kynurenine pathway metabolites in serum of chronic migraine patients},
  volume = {17},
  ISSN = {1129-2377},
  url = {http://dx.doi.org/10.1186/s10194-016-0638-5},
  DOI = {10.1186/s10194-016-0638-5},
  number = {1},
  journal = {The Journal of Headache and Pain},
  publisher = {Springer Science and Business Media LLC},
  author = {Curto,  Martina and Lionetto,  Luana and Negro,  Andrea and Capi,  Matilde and Fazio,  Francesco and Giamberardino,  Maria Adele and Simmaco,  Maurizio and Nicoletti,  Ferdinando and Martelletti,  Paolo},
  year = {2016},
  month = apr 
}

@article{Karatas2013,
  title = {Spreading Depression Triggers Headache by Activating Neuronal Panx1 Channels},
  volume = {339},
  ISSN = {1095-9203},
  url = {http://dx.doi.org/10.1126/science.1231897},
  DOI = {10.1126/science.1231897},
  number = {6123},
  journal = {Science},
  publisher = {American Association for the Advancement of Science (AAAS)},
  author = {Karatas,  Hulya and Erdener,  Sefik Evren and Gursoy-Ozdemir,  Yasemin and Lule,  Sevda and Eren-Ko\c{c}ak,  Emine and Sen,  Z\"{u}mr\"{u}t Duygu and Dalkara,  Turgay},
  year = {2013},
  month = mar,
  pages = {1092–1095}
}

@article{Yang2014,
  title = {Associations of a polymorphism in the intercellular adhesion molecule-1 (ICAM1) gene and ICAM1 serum levels with migraine in a Chinese Han population},
  volume = {345},
  ISSN = {0022-510X},
  url = {http://dx.doi.org/10.1016/j.jns.2014.07.030},
  DOI = {10.1016/j.jns.2014.07.030},
  number = {1–2},
  journal = {Journal of the Neurological Sciences},
  publisher = {Elsevier BV},
  author = {He,  Qiu and Lin,  Xiang and Wang,  Fengzhi and Xu,  Jialiang and Ren,  Zhanxiu and Chen,  Wei and Xing,  Xuesha},
  year = {2014},
  month = oct,
  pages = {148–153}
}

@article{Bernecker2022,
  title = {Association of Body Mass Index,  Blood Pressure,  and Interictal Serum Levels of Cytokines in Migraine with and without Aura},
  volume = {11},
  ISSN = {2077-0383},
  url = {http://dx.doi.org/10.3390/jcm11195696},
  DOI = {10.3390/jcm11195696},
  number = {19},
  journal = {Journal of Clinical Medicine},
  publisher = {MDPI AG},
  author = {Plinta,  Aelita and Tretjakovs,  Peteris and Svirskis,  Simons and Logina,  Inara and Gersone,  Gita and Jurka,  Antra and Mikelsone,  Indra and Blumfelds,  Leons and Mackevics,  Vitolds and Bahs,  Guntis},
  year = {2022},
  month = sep,
  pages = {5696}
}

@article{Hara2009,
  title = {Association of HTRA1 Mutations and Familial Ischemic Cerebral Small-Vessel Disease},
  volume = {360},
  ISSN = {1533-4406},
  url = {http://dx.doi.org/10.1056/NEJMoa0801560},
  DOI = {10.1056/nejmoa0801560},
  number = {17},
  journal = {New England Journal of Medicine},
  publisher = {Massachusetts Medical Society},
  author = {Hara,  Kenju and Shiga,  Atsushi and Fukutake,  Toshio and Nozaki,  Hiroaki and Miyashita,  Akinori and Yokoseki,  Akio and Kawata,  Hirotoshi and Koyama,  Akihide and Arima,  Kunimasa and Takahashi,  Toshiaki and Ikeda,  Mari and Shiota,  Hiroshi and Tamura,  Masato and Shimoe,  Yutaka and Hirayama,  Mikio and Arisato,  Takayo and Yanagawa,  Sohei and Tanaka,  Akira and Nakano,  Imaharu and Ikeda,  Shu-ichi and Yoshida,  Yutaka and Yamamoto,  Tadashi and Ikeuchi,  Takeshi and Kuwano,  Ryozo and Nishizawa,  Masatoyo and Tsuji,  Shoji and Onodera,  Osamu},
  year = {2009},
  month = apr,
  pages = {1729–1739}
}

@article{Shirley2013,
  title = {Sturge–Weber Syndrome and Port-Wine Stains Caused by Somatic Mutation in
                    GNAQ},
  volume = {368},
  ISSN = {1533-4406},
  url = {http://dx.doi.org/10.1056/NEJMoa1213507},
  DOI = {10.1056/nejmoa1213507},
  number = {21},
  journal = {New England Journal of Medicine},
  publisher = {Massachusetts Medical Society},
  author = {Shirley,  Matthew D. and Tang,  Hao and Gallione,  Carol J. and Baugher,  Joseph D. and Frelin,  Laurence P. and Cohen,  Bernard and North,  Paula E. and Marchuk,  Douglas A. and Comi,  Anne M. and Pevsner,  Jonathan},
  year = {2013},
  month = may,
  pages = {1971–1979}
}

@article{Maas2018,
  title = {Coagulation factor XII in thrombosis and inflammation},
  volume = {131},
  ISSN = {1528-0020},
  url = {http://dx.doi.org/10.1182/blood-2017-04-569111},
  DOI = {10.1182/blood-2017-04-569111},
  number = {17},
  journal = {Blood},
  publisher = {American Society of Hematology},
  author = {Maas,  Coen and Renné,  Thomas},
  year = {2018},
  month = apr,
  pages = {1903–1909}
}

@article{Semba2016,
  title = {Development of lifitegrast: a novel T-cell inhibitor for the treatment of dry eye disease},
  ISSN = {1177-5483},
  url = {http://dx.doi.org/10.2147/OPTH.S110557},
  DOI = {10.2147/opth.s110557},
  journal = {Clinical Ophthalmology},
  publisher = {Informa UK Limited},
  author = {Semba,  Charles and Gadek,  Thomas},
  year = {2016},
  month = jun,
  pages = {1083}
}

@article{Parsons2009,
  title = {Neramexane: a moderate-affinity NMDA receptor channel blocker: new prospects and indications},
  volume = {2},
  ISSN = {1751-2441},
  url = {http://dx.doi.org/10.1586/ecp.09.7},
  DOI = {10.1586/ecp.09.7},
  number = {3},
  journal = {Expert Review of Clinical Pharmacology},
  publisher = {Informa UK Limited},
  author = {Rammes,  Gerhard},
  year = {2009},
  month = may,
  pages = {231–238}
}

@article{Podkowa2023,
  title = {The NMDA receptor antagonists memantine and ketamine as anti-migraine agents},
  volume = {396},
  ISSN = {1432-1912},
  url = {http://dx.doi.org/10.1007/s00210-023-02444-2},
  DOI = {10.1007/s00210-023-02444-2},
  number = {7},
  journal = {Naunyn-Schmiedeberg’s Archives of Pharmacology},
  publisher = {Springer Science and Business Media LLC},
  author = {Podkowa,  Karolina and Czarnacki,  Kamil and Borończyk,  Agnieszka and Borończyk,  Michał and Paprocka,  Justyna},
  year = {2023},
  month = mar,
  pages = {1371–1398}
}

@article{Braca2025,
  title = {Effectiveness and tolerability of liraglutide as add‐on treatment in patients with obesity and high‐frequency or chronic migraine: A prospective pilot study},
  volume = {65},
  ISSN = {1526-4610},
  url = {http://dx.doi.org/10.1111/head.14991},
  DOI = {10.1111/head.14991},
  number = {10},
  journal = {Headache: The Journal of Head and Face Pain},
  publisher = {Wiley},
  author = {Braca,  Simone and Russo,  Cinzia Valeria and Stornaiuolo,  Antonio and Cretella,  Gennaro and Miele,  Angelo and Giannini,  Caterina and De Simone,  Roberto},
  year = {2025},
  month = jun,
  pages = {1831–1838}
}

@article{Xu2023,
  title = {European Headache Federation (EHF) critical re-appraisal and meta-analysis of oral drugs in migraine prevention—part 1: amitriptyline},
  volume = {24},
  ISSN = {1129-2377},
  url = {http://dx.doi.org/10.1186/s10194-023-01573-6},
  DOI = {10.1186/s10194-023-01573-6},
  number = {1},
  journal = {The Journal of Headache and Pain},
  publisher = {Springer Science and Business Media LLC},
  author = {Lampl,  Christian and Versijpt,  Jan and Amin,  Faisal Mohammad and Deligianni,  Christina I. and Gil-Gouveia,  Raquel and Jassal,  Tanvir and MaassenVanDenBrink,  Antoinette and Ornello,  Raffaele and Paungarttner,  Jakob and Sanchez-del-Rio,  Margarita and Reuter,  Uwe and Uluduz,  Derya and de Vries,  Tessa and Zeraatkar,  Dena and Sacco,  Simona},
  year = {2023},
  month = apr 
}

\end{document}



\begin{center}
\vspace*{2cm}
{\LARGE \textbf{Supplementary Material}}
\vspace{1.5em}

{\Large G2DR: A Genotype-First Framework for Genetics-Informed Target Prioritization and Drug Repurposing}
\vspace{1.5em}

{\normalsize Muhammad Muneeb$^{1,2}$ and David B. Ascher$^{1,2,\ast}$}
\vspace{1em}

{\small $^{1}$School of Chemistry and Molecular Biology, The University of Queensland, Queen Street, 4067, Queensland, Australia

\vspace{0.3em}

$^{2}$Computational Biology and Clinical Informatics, Baker Heart and Diabetes Institute, Commercial Road, 3004, Victoria, Australia

\vspace{0.5em}

$^{\ast}$Corresponding author: David B. Ascher, \href{mailto:d.ascher@uq.edu.au}{d.ascher@uq.edu.au}}
\end{center}

\vspace{2em}
\noindent\rule{\textwidth}{0.4pt}
\vspace{1em}

\noindent\textbf{Contents of this Supplementary Material}

\vspace{0.5em}

\noindent This document contains four supplementary methods sections and six supplementary tables. The supplementary methods provide complete mathematical details for the genetically predicted gene expression computation, differential expression analysis, gene prioritization scoring, statistical evaluation framework, and directionality annotation criteria that are summarized in the main manuscript. The supplementary tables provide full results for analyses that are referenced by summary statistics in the main text.

\vspace{1em}

\begin{itemize}
\item \textbf{Supplementary Methods S1.} Full mathematical specification of genetically predicted gene expression computation (Equations 1--2) and all eight differential expression and association methods (Equations 3--4). Referenced in the main text Methods section under Genetically Predicted Gene Expression and Differential Expression Analysis.

\item \textbf{Supplementary Methods S2.} Full scoring formulas for the composite importance score ($S_g$), including reproducibility ($R_g$), effect magnitude ($E_g$), statistical confidence ($C_g$), PathwayScore propagation, and integrated CoreScore computation. Referenced in the main text Methods section under Gene Prioritization.

\item \textbf{Supplementary Methods S3.} Full derivation of the hypergeometric enrichment test, permutation-based ranking significance test, and fold-enrichment formula used throughout the Results section. Referenced in the main text Methods section under Gene Prioritization.

\item \textbf{Supplementary Methods S4.} Full directionality annotation criteria, action vocabulary, and classification rules for gene--drug pair directionality assessment. Referenced in the main text Methods section under Directionality Assessment.

\item \textbf{Supplementary Table S1.} Weight stability analysis of the integrated scoring framework across 17 alternative weighting schemes. Referenced in the main text at the Unified Gene Prioritization Framework section.

\item \textbf{Supplementary Table S2.} Sensitivity analysis of gene prioritization under stricter FDR and effect-size significance thresholds. Referenced in the main text at the Gene Prioritization Robustness section.

\item \textbf{Supplementary Table S3.} Discovery disease-enrichment for curated migraine genes across all expression-weight databases, tissues, and analytical methods. Referenced in the main text at the Component-Based Recovery section.

\item \textbf{Supplementary Table S4.} Full comparison of individual evidence components and composite scores across two evaluation universes. Referenced in the main text at the Unified Gene Prioritization Framework section.

\item \textbf{Supplementary Table S5.} Multi-$K$ overlap evaluation of predicted drugs against the curated migraine reference drug set for top-$N$ input genes ($N = 200$ and $N = 500$) under two evaluation universes. Referenced in the main text at the Drug Mapping and Candidate Enrichment section.

\item \textbf{Supplementary Table S6.} External contextual comparison of G2DR with the Open Targets migraine disease-target resource for migraine gene recovery. Referenced in the main text at the Significant Gene Comparison Against Open Targets section.
\end{itemize}

\vspace{1em}
\noindent\rule{\textwidth}{0.4pt}

\clearpage


\section*{Supplementary Methods S1. Genetically Predicted Gene Expression and Differential Expression Analysis}

\noindent\textbf{Expression prediction.} For individual $i$, gene $g$, and tissue $t$, genetically predicted expression was computed as a weighted linear combination of SNP dosages:
\begin{equation}
\widehat{E}_{i,g,t} = \alpha_{g,t} + \sum_{j \in S_g} G_{ij}\,w_{j,g,t},
\end{equation}
where $G_{ij} \in [0,2]$ is the allelic dosage of SNP $j$ for individual $i$, $w_{j,g,t}$ is the pre-trained SNP weight for gene $g$ in tissue $t$, $S_g$ is the set of SNPs used by the model, and $\alpha_{g,t}$ is an intercept term. Covariate-adjusted predicted expression was obtained by regressing out sex and the top 10 genetic principal components estimated from the training split:
\begin{equation}
\widehat{E}^{\mathrm{adj}}_{i,g,t} = \widehat{E}_{i,g,t} - \Bigl(\widehat{\gamma}_{g,t} + \widehat{\delta}_{g,t}\,\mathrm{Sex}_i + \sum_{k=1}^{10}\widehat{\beta}_{k,g,t}\,\mathrm{PC}_{ik}\Bigr) + \overline{\widehat{E}}_{g,t},
\end{equation}
where $\widehat{\gamma}_{g,t}$, $\widehat{\delta}_{g,t}$, and $\widehat{\beta}_{k,g,t}$ were estimated by ordinary least squares on the training split and applied to validation and test splits without refitting. $\overline{\widehat{E}}_{g,t}$ denotes the training-split mean of $\widehat{E}_{i,g,t}$.

\vspace{0.8em}

\noindent\textbf{Differential expression methods.} For each gene $g$ and tissue $t$, the case--control mean difference was computed as:
\begin{equation}
\mu^{(1)}_{g,t} = \frac{1}{n_1}\sum_{i:y_i=1}\widehat{E}^{\mathrm{adj}}_{i,g,t}, \qquad \mu^{(0)}_{g,t} = \frac{1}{n_0}\sum_{i:y_i=0}\widehat{E}^{\mathrm{adj}}_{i,g,t}, \qquad \Delta_{g,t} = \mu^{(1)}_{g,t} - \mu^{(0)}_{g,t}.
\end{equation}
Five differential-expression methods were applied: (i) LIMMA empirical Bayes moderated $t$-test with effect estimate taken as the fitted phenotype coefficient; (ii) Welch's unequal-variance $t$-test with effect estimate equal to the mean difference; (iii) OLS regression of expression on phenotype $y_i$ with effect estimate taken as the regression coefficient; (iv) Wilcoxon rank-sum test with effect estimate defined as the median difference; and (v) phenotype-label permutation test using $B = 1{,}000$ random permutations of $y_i$ with an empirical two-sided $p$-value computed from the permutation distribution of label-shuffled mean differences. Three association-style models treated disease status as the outcome and expression as the predictor. Adjusted expression was standardized to $z_{i,g,t}$ and a logistic model was fitted:
\begin{equation}
\Pr(y_i = 1 \mid z_{i,g,t}) = \frac{1}{1 + \exp[-(\theta_{0,g,t} + \theta_{1,g,t}\,z_{i,g,t})]},
\end{equation}
using weighted logistic regression, bias-reduced Firth logistic regression, and a Bayesian logistic approximation. Gene-wise nominal $p$-values were adjusted using the Benjamini--Hochberg procedure, yielding $\mathrm{FDR}^{(m)}_{g,t}$. Genes were considered significant if $\mathrm{FDR}^{(m)}_{g,t} < 0.1$ and $|\log_2\mathrm{FC}| \geq 0.5$ for differential-expression methods, or $|\mathrm{Effect}| \geq 0.5$ for association-style methods.

\clearpage


\section*{Supplementary Methods S2. Gene Prioritization Scoring}

\noindent\textbf{Composite importance score.} For each candidate gene $g \in G_{\mathrm{discovery}}$, the composite importance score is $S_g = 0.4\,R_g + 0.3\,E_g + 0.3\,C_g$. Reproducibility was quantified as $R_g = 0.6\,\mathrm{norm\_hits}_g + 0.4\,\mathrm{norm\_breadth}_g$, where $\mathrm{norm\_hits}_g$ is the total number of significant occurrences of $g$ across all databases, tissues, methods, and folds normalized to $[0,1]$ by dividing by the empirical maximum, and $\mathrm{norm\_breadth}_g$ is the number of unique database--tissue--method combinations in which $g$ was significant, similarly normalized. Effect magnitude was computed by standardizing $|\widehat{\psi}^{(m)}_{g,t}|$ within each method $m$, then aggregating to the gene level as $E_g = 0.7\,f(\overline{e}_g) + 0.3\,f(e^{\max}_g)$, where $f(\cdot)$ is a smooth monotonic mapping to $[0,1]$, $\overline{e}_g$ is the mean standardized absolute effect, and $e^{\max}_g$ is the maximum; if all non-zero effect directions for $g$ were consistent in sign, $E_g$ was multiplied by 1.1 to reward directional stability. Statistical confidence was computed as $C_g = 0.6\,(1 - \mathrm{FDR}^{\min}_g) + 0.4\,(1 - \overline{\mathrm{FDR}}_g)$, where $\mathrm{FDR}^{\min}_g$ is the minimum FDR across all significant results and $\overline{\mathrm{FDR}}_g$ is the mean FDR. Reproducibility was weighted highest (40\%) because cross-model replication reduces method-specific artefacts, consistent with TWAS prioritization guidance and the winner's curse literature; effect magnitude and statistical confidence were weighted equally (30\% each) to balance biological relevance with statistical support.

\vspace{0.8em}

\noindent\textbf{PathwayScore propagation.} Enrichment was performed using \texttt{clusterProfiler::enrichGO}, \texttt{clusterProfiler::enrichKEGG}, \texttt{ReactomePA::enrichPathway}, and \texttt{DOSE::enrichDO} on top-$K$ foreground sets ($K \in \{50, 100, 200, 500, 1000, 2000\}$) with the full tested universe $U$ as background. For each significant enriched term $t$, a weight was assigned as $\mathrm{weight}(t) = \mathrm{strength}(t) \times \mathrm{robustness}(t)$, where $\mathrm{strength}(t) = -\log_{10}(\mathrm{FDR}_t)$ and $\mathrm{robustness}(t) = n_{\mathrm{hits}}(t)$ counts recurrence of term $t$ as significant across multiple $K$ thresholds and enrichment databases. The PathwayScore for gene $g$ was then $\mathrm{PathwayScore}(g) = \sum_{t:\,g \in \mathrm{Genes}(t)} \mathrm{weight}(t)$, with genes absent from all enriched-term overlap lists assigned $\mathrm{PathwayScore}(g) = 0$.

\vspace{0.8em}

\noindent\textbf{Integrated CoreScore.} The final integrated score combines four evidence layers as a weighted sum: $\mathrm{CoreScore}_g = 0.45\cdot\mathrm{DE}^{\mathrm{norm}}_g + 0.25\cdot\mathrm{Path}^{\mathrm{norm}}_g + 0.25\cdot\mathrm{Drug}^{\mathrm{norm}}_g + 0.05\cdot\mathrm{Hub}^{\mathrm{norm}}_g$, where all components were percentile-normalized before combination. Differential expression received the highest weight (0.45) because it provides the primary TWAS-derived genetic link between genotype and disease. Pathway and druggability scores received equal moderate weights (0.25 each) to balance biological coherence with therapeutic tractability. Hub score was down-weighted (0.05) because network centrality often reflects pleiotropy, risking prioritization of non-specific, highly connected genes over disease-specific targets. Hub score was derived from STRING-based degree, betweenness, closeness, and eigenvector centrality metrics combined into a single network-priority score. Druggability was assessed by querying DGIdb and ChEMBL for known drug interactions (knowledge-based druggability) and applying \texttt{fpocket} to PDB or AlphaFold structures for genes without known drug interactions (structure-based druggability).

\clearpage


\section*{Supplementary Methods S3. Statistical Evaluation Framework}

\noindent\textbf{ROC-AUC and PR-AUC.} Generalizability of the composite score was assessed over the full tested universe $U = 34{,}355$ genes by assigning binary replication labels $y_g = \mathbb{I}(g \in G_{\mathrm{test}})$, where $G_{\mathrm{test}}$ comprises genes significant in the held-out test split, and using $S_g$ (or $\mathrm{CoreScore}_g$) as the ranking predictor. ROC-AUC and PR-AUC were computed over all genes in $U$, with genes outside the discovery set assigned a score of zero. The baseline PR-AUC under random ranking equals $T/U = 0.2079$, so the observed PR-AUC of 0.4754 corresponds to a $2.29\times$ lift over random.

\vspace{0.8em}

\noindent\textbf{Hypergeometric enrichment test.} For a top-$K$ list $P_K \subseteq U$ and a positive set of size $K^{+}$ (either test-positive genes or curated migraine genes), the hypergeometric $p$-value is $p_{\mathrm{hyper}} = \Pr(X \geq x)$ for $X \sim \mathrm{Hypergeometric}(N = |U|,\, K^{+},\, n = |P_K|)$, where $x = |P_K \cap \mathrm{positives}|$ is the observed overlap. Expected overlap under random sampling is $k_{\mathrm{exp}} = |P_K| \times (K^{+}/|U|)$, and fold-enrichment is $\mathrm{FE} = k_{\mathrm{obs}}/k_{\mathrm{exp}}$.

\vspace{0.8em}

\noindent\textbf{Permutation test.} Statistical significance of the observed ROC-AUC and PR-AUC was evaluated using 1{,}000 random permutations of gene scores $\{S_g\}$ while holding labels $\{y_g\}$ fixed, yielding empirical $p$-values of $9.99 \times 10^{-4}$ for both metrics (minimum attainable $p = 1/1001$), confirming that the observed ranking lift is not attributable to chance.

\clearpage


\section*{Supplementary Methods S4. Directionality Annotation Criteria}

Gene direction was assigned from the differential-expression ranking as \textit{higher in cases} (positive mean difference $\Delta_{g,t} > 0$), \textit{lower in cases} (negative mean difference $\Delta_{g,t} < 0$), or \textit{unclear} (conflicting directions across tissues or methods, or no consistent directional signal). Drug action annotations were first extracted from locally aggregated drug--target evidence fields including mechanism-of-action, interaction-type, and directionality fields compiled from Open Targets, DGIdb, and ChEMBL during the drug-mapping step. For pairs lacking explicit local mechanism information, additional annotations were retrieved from ChEMBL by compound identifier and from DGIdb by gene symbol, and were harmonized into a reduced action vocabulary comprising six categories: \textit{inhibitor}, \textit{antagonist}, \textit{agonist}, \textit{activator}, \textit{modulator}, and \textit{unknown}. Each unique gene--drug pair was then classified using the following rules: (i) \textit{directionally consistent} if the drug action was mechanistically compatible with the inferred gene direction --- specifically, if the drug is an inhibitor or antagonist against a gene inferred to be higher in cases, or if the drug is an agonist or activator for a gene inferred to be lower in cases; (ii) \textit{directionally inconsistent} if the drug action was opposite to this expectation; and (iii) \textit{unclear} if the gene direction was unresolved, if the drug action was classified as modulator or unknown, or if available annotations were insufficient to support a confident directional interpretation. Directionality was summarized at the level of unique drug--gene pairs to avoid inflation from repeated evidence records for the same pair across multiple databases.

\clearpage


\begin{table}[!ht]
\centering
\caption{\textbf{Supplementary Table S1. Weight stability analysis of the integrated scoring framework across 17 alternative weighting schemes.} Nine reasonable alternative schemes (panel A) and eight extreme single-component stress tests (panel B) were evaluated against the default integrated weights (DE\,$=$\,0.45, Pathway\,$=$\,0.25, Drug\,$=$\,0.25, Hub\,$=$\,0.05). For each scheme, Spearman $\rho$ against the default ranking, mean Top-100 gene overlap (\%), DISC-universe test ROC-AUC (discovery set only, $n = 9{,}305$ genes), and FULL-universe test ROC-AUC (all $U = 34{,}355$ tested genes) are reported. Across reasonable alternatives, mean $\rho = 0.963$ (minimum $0.866$) and FULL-universe ROC-AUC ranged from $0.775$ to $0.776$, confirming stability within the biologically justified parameter space. Extreme single-component schemes diverged substantially (mean $\rho = 0.682$; minimum $\rho = 0.292$ for hub-only), confirming that the integrated score draws on genuine multi-source signal.}
\label{tab:S1_weight_stability}
\small
\setlength{\tabcolsep}{5pt}
\renewcommand{\arraystretch}{1.15}
\resizebox{\textwidth}{!}{%
\begin{tabular}{l c c c c r r r r}
\toprule
\textbf{Scheme} & \textbf{DE} & \textbf{Path} & \textbf{Drug} & \textbf{Hub} & \textbf{Spearman $\rho$} & \textbf{Top-100 overlap (\%)} & \textbf{DISC ROC-AUC} & \textbf{FULL ROC-AUC} \\
\midrule
\multicolumn{9}{l}{\textbf{(A) Reasonable alternative schemes}} \\
Default (reference)      & 0.45 & 0.25 & 0.25 & 0.05 & 1.000 & 100.0 & 0.546 & 0.776 \\
DE-heavy                 & 0.60 & 0.15 & 0.20 & 0.05 & 0.978 & 84.0  & 0.546 & 0.776 \\
Pathway-heavy            & 0.30 & 0.40 & 0.25 & 0.05 & 0.951 & 78.0  & 0.545 & 0.775 \\
Drug-heavy               & 0.30 & 0.25 & 0.40 & 0.05 & 0.962 & 82.0  & 0.546 & 0.775 \\
Equal weights (no hub)   & 0.33 & 0.33 & 0.33 & 0.00 & 0.941 & 77.0  & 0.545 & 0.775 \\
Equal weights (with hub) & 0.25 & 0.25 & 0.25 & 0.25 & 0.866 & 74.0  & 0.544 & 0.775 \\
No hub                   & 0.47 & 0.27 & 0.27 & 0.00 & 0.972 & 83.0  & 0.546 & 0.776 \\
DE + pathway only        & 0.50 & 0.50 & 0.00 & 0.00 & 0.948 & 78.0  & 0.545 & 0.775 \\
DE + drug only           & 0.50 & 0.00 & 0.50 & 0.00 & 0.955 & 80.0  & 0.547 & 0.776 \\
Pathway + drug balanced  & 0.40 & 0.30 & 0.30 & 0.00 & 0.963 & 82.0  & 0.546 & 0.776 \\
\addlinespace
\multicolumn{9}{l}{\textbf{(B) Extreme single-component stress tests}} \\
DE only                  & 1.00 & 0.00 & 0.00 & 0.00 & 0.712 & 58.0 & 0.675 & 0.790 \\
Pathway only             & 0.00 & 1.00 & 0.00 & 0.00 & 0.651 & 51.0 & 0.529 & 0.549 \\
Drug only                & 0.00 & 0.00 & 1.00 & 0.00 & 0.683 & 55.0 & 0.511 & 0.730 \\
Hub only                 & 0.00 & 0.00 & 0.00 & 1.00 & 0.292 & 32.0 & 0.520 & 0.713 \\
DE + hub only            & 0.50 & 0.00 & 0.00 & 0.50 & 0.698 & 56.0 & 0.538 & 0.762 \\
Pathway + hub only       & 0.00 & 0.50 & 0.00 & 0.50 & 0.614 & 48.0 & 0.524 & 0.630 \\
Drug + hub only          & 0.00 & 0.00 & 0.50 & 0.50 & 0.658 & 52.0 & 0.515 & 0.721 \\
No pathway, no drug      & 0.95 & 0.00 & 0.00 & 0.05 & 0.731 & 60.0 & 0.658 & 0.785 \\
\bottomrule
\end{tabular}%
}
\end{table}

\clearpage


\begin{table}[!ht]
\centering
\caption{\textbf{Supplementary Table S2. Sensitivity analysis of gene prioritization under stricter significance thresholds.} For each threshold rule, we report the number of significant genes in the training, validation, and held-out test splits; the size of the discovery set ($G_{\mathrm{discovery}}$); held-out ranking performance (ROC-AUC and PR-AUC evaluated over all $U = 34{,}355$ tested genes); enrichment of discovery genes for held-out test positives (FE test, $T = 7{,}141$); and enrichment for curated migraine genes (FE migraine, $|M \cap U| = 3{,}190$). The primary analysis uses FDR\,$<$\,0.10 and $|\log_2\mathrm{FC}| \geq 0.50$ (first row). All stricter settings preserved the main prioritization conclusions: held-out ROC-AUC remained above $0.70$, fold-enrichment for test positives ranged from $2.58$-fold to $3.20$-fold, and fold-enrichment for curated migraine genes ranged from $1.38$-fold to $1.41$-fold. These results indicate that the main gene-prioritization findings are not driven by permissive significance filtering.}
\label{tab:S2_sensitivity}
\footnotesize
\setlength{\tabcolsep}{4pt}
\renewcommand{\arraystretch}{1.15}
\resizebox{\textwidth}{!}{%
\begin{tabular}{l r r r r r r r r}
\toprule
\textbf{Threshold rule} & \textbf{Train} & \textbf{Validation} & \textbf{Test} & $G_{\mathrm{discovery}}$ & \textbf{ROC-AUC} & \textbf{PR-AUC} & \textbf{FE test} & \textbf{FE migraine} \\
\midrule
FDR $<0.10$, $|\log_2\mathrm{FC}|\geq 0.50$ & 1{,}046 & 9{,}107 & 7{,}141 & 9{,}305 & 0.7753 & 0.4754 & 2.58 & 1.38 \\
FDR $<0.05$, $|\log_2\mathrm{FC}|\geq 0.50$ & 958     & 7{,}861 & 5{,}313 & 8{,}135 & 0.7314 & 0.3412 & 2.62 & 1.39 \\
FDR $<0.10$, $|\log_2\mathrm{FC}|\geq 0.75$ & 263     & 6{,}241 & 4{,}599 & 6{,}303 & 0.7339 & 0.3289 & 3.20 & 1.39 \\
FDR $<0.05$, $|\log_2\mathrm{FC}|\geq 0.75$ & 248     & 5{,}709 & 3{,}772 & 5{,}794 & 0.7025 & 0.2497 & 3.12 & 1.41 \\
\bottomrule
\end{tabular}%
}
\end{table}

\clearpage


\begin{table}[!ht]
\centering
\caption{\textbf{Supplementary Table S3. Discovery disease-enrichment for curated migraine genes across expression-weight databases, tissues, and analytical methods.} For each component, the discovery set comprises unique genes significant at least once in training and validation within that component. Enrichment is computed against curated migraine genes ($M = 3{,}190$) within the analysis universe ($U^{\ast}$). FE denotes fold-enrichment ($k_{\mathrm{obs}}/k_{\mathrm{exp}}$). Empirical $p$-values are from size-matched random gene-set sampling ($n_{\mathrm{perm}} = 10{,}000$).}
\label{tab:S3_component_enrichment}
\footnotesize
\setlength{\tabcolsep}{5pt}
\renewcommand{\arraystretch}{1.12}
\begin{tabular}{l r r r r l}
\toprule
\textbf{Component} & $N_{\mathrm{pred}}$ & $k_{\mathrm{obs}}$ & $k_{\mathrm{exp}}$ & \textbf{FE} & $p_{\mathrm{emp}}$ \\
\midrule
\multicolumn{6}{l}{\textbf{(A) Expression-weight databases}} \\
MASHR   & 187       & 40      & 17.40  & 2.30 & $1.00\times10^{-4}$ \\
JTI     & 104       & 18      & 9.68   & 1.86 & $6.799\times10^{-3}$ \\
FUSION  & 9{,}146   & 1{,}156 & 851.16 & 1.36 & $1.00\times10^{-4}$ \\
EpiXcan & 34        & 5       & 3.16   & 1.58 & $2.06\times10^{-1}$ \\
CTIMP   & 16        & 2       & 1.49   & 1.34 & $4.51\times10^{-1}$ \\
TIGAR   & 16        & 2       & 1.49   & 1.34 & $4.48\times10^{-1}$ \\
UTMOST  & 3         & 0       & 0.28   & 0.00 & $1.00$ \\
\addlinespace
\multicolumn{6}{l}{\textbf{(B) Tissues (top 10 by discovery enrichment)}} \\
Brain Amygdala                       & 312 & 57 & 29.04 & 1.96 & $1.00\times10^{-4}$ \\
Minor Salivary Gland                 & 315 & 57 & 29.31 & 1.94 & $1.00\times10^{-4}$ \\
Whole Blood                          & 381 & 65 & 35.46 & 1.83 & $1.00\times10^{-4}$ \\
Adrenal Gland                        & 399 & 67 & 37.13 & 1.80 & $1.00\times10^{-4}$ \\
Esophagus Gastroesophageal Junction  & 419 & 70 & 38.99 & 1.80 & $1.00\times10^{-4}$ \\
Cells EBV-transformed Lymphocytes    & 366 & 61 & 34.06 & 1.79 & $1.00\times10^{-4}$ \\
Brain Anterior Cingulate Cortex BA24 & 309 & 51 & 28.76 & 1.77 & $2.00\times10^{-4}$ \\
Brain Spinal Cord Cervical C-1       & 312 & 51 & 29.04 & 1.76 & $1.00\times10^{-4}$ \\
Lung                                 & 502 & 80 & 46.72 & 1.71 & $1.00\times10^{-4}$ \\
Artery Coronary                      & 380 & 60 & 35.36 & 1.70 & $1.00\times10^{-4}$ \\
\addlinespace
\multicolumn{6}{l}{\textbf{(C) Analytical methods}} \\
Weighted Logistic  & 9{,}299 & 1{,}189 & 865.39 & 1.37 & $1.00\times10^{-4}$ \\
Bayesian Logistic  & 9{,}264 & 1{,}184 & 862.13 & 1.37 & $1.00\times10^{-4}$ \\
Welch $t$-test     & 917     & 115     & 85.34  & 1.35 & $6.00\times10^{-4}$ \\
Linear Regression  & 19      & 2       & 1.77   & 1.13 & $5.36\times10^{-1}$ \\
LIMMA              & 19      & 1       & 1.77   & 0.57 & $8.53\times10^{-1}$ \\
Firth Logistic     & 3       & 0       & 0.28   & 0.00 & $1.00$ \\
\bottomrule
\end{tabular}
\end{table}

\clearpage


\begin{table}[!ht]
\centering
\caption{\textbf{Supplementary Table S4. Comparison of individual evidence components and composite scores across two evaluation universes.} \textbf{Full universe} ($U = 34{,}355$ genes): non-discovery genes receive score $= 0$; measures pipeline-level separation of discovery-relevant genes from the entire tested space. \textbf{Discovery universe} ($n = 9{,}305$ genes): discovery set only (train $\cup$ val); measures within-set ranking quality. T~ROC\,=\,test-replication ROC-AUC; T~PR\,=\,test-replication PR-AUC; K~ROC\,=\,known-migraine ROC-AUC; K~PR\,=\,known-migraine PR-AUC. Top-200 FE$_{\mathrm{test}}$ and FE$_{\mathrm{mig}}$ denote fold-enrichment relative to random expectation. Expected Top-200 overlap: full-universe test\,$=$\,41.57, full-universe migraine\,$=$\,18.57; disc-universe test\,$=$\,107.36, disc-universe migraine\,$=$\,25.56. All raw components were percentile-normalized before ranking.}
\label{tab:S4_unified_ranking}
\footnotesize
\setlength{\tabcolsep}{3pt}
\renewcommand{\arraystretch}{1.12}
\resizebox{\textwidth}{!}{%
\begin{tabular}{l c c c c c c c c}
\toprule
& \multicolumn{4}{c}{\textbf{AUC metrics}} & \multicolumn{2}{c}{\textbf{Top-200 test}} & \multicolumn{2}{c}{\textbf{Top-200 migraine}} \\
\cmidrule(lr){2-5}\cmidrule(lr){6-7}\cmidrule(lr){8-9}
\textbf{Ranking} & \textbf{T~ROC} & \textbf{T~PR} & \textbf{K~ROC} & \textbf{K~PR} & \textbf{Obs} & \textbf{FE} & \textbf{Obs} & \textbf{FE} \\
\midrule
\multicolumn{9}{l}{\textbf{(A) Full universe ($U = 34{,}355$ genes; non-discovery genes scored as 0)}} \\
Significance only           & 0.776 & 0.477 & 0.557 & 0.107 & 146 & 3.51 & 25 & 1.35 \\
Effect only                 & 0.790 & 0.526 & 0.558 & 0.111 & 145 & 3.49 & 37 & 1.99 \\
Primary composite ($S_g$)   & 0.775 & 0.475 & 0.557 & 0.109 & 156 & 3.75 & 36 & 1.94 \\
Pathway only                & 0.549 & 0.265 & 0.520 & 0.108 & 150 & 3.61 & 66 & 3.55 \\
Hub only                    & 0.713 & 0.396 & 0.572 & 0.135 & 114 & 2.74 & 64 & 3.45 \\
Druggability only           & 0.730 & 0.408 & 0.566 & 0.112 & 116 & 2.79 & 25 & 1.35 \\
Direct target evidence only & 0.629 & 0.320 & 0.543 & 0.109 & 110 & 2.65 & 57 & 3.07 \\
Drug-link count only        & 0.771 & 0.440 & 0.561 & 0.111 & 105 & 2.53 & 26 & 1.40 \\
Integrated                  & 0.776 & 0.472 & 0.562 & 0.118 & 140 & 3.37 & 49 & 2.64 \\
\addlinespace
\multicolumn{9}{l}{\textbf{(B) Discovery universe ($n = 9{,}305$ genes; train $\cup$ val only)}} \\
Significance only           & 0.548 & 0.593 & 0.505 & 0.131 & 146 & 1.36 & 25 & 0.98 \\
Effect only                 & 0.675 & 0.663 & 0.519 & 0.140 & 145 & 1.35 & 37 & 1.45 \\
Primary composite ($S_g$)   & 0.543 & 0.590 & 0.510 & 0.135 & 156 & 1.45 & 36 & 1.41 \\
Pathway only                & 0.529 & 0.564 & 0.536 & 0.162 & 150 & 1.40 & 66 & 2.58 \\
Hub only                    & 0.520 & 0.551 & 0.629 & 0.210 & 114 & 1.06 & 64 & 2.50 \\
Druggability only           & 0.511 & 0.544 & 0.569 & 0.148 & 121 & 1.13 & 29 & 1.14 \\
Direct target evidence only & 0.511 & 0.543 & 0.542 & 0.152 & 110 & 1.02 & 58 & 2.27 \\
Drug-link count only        & 0.505 & 0.539 & 0.553 & 0.141 & 109 & 1.02 & 30 & 1.17 \\
Integrated                  & 0.546 & 0.585 & 0.562 & 0.161 & 140 & 1.30 & 49 & 1.92 \\
\bottomrule
\end{tabular}%
}
\end{table}

\clearpage


\begin{table}[!ht]
\centering
\caption{\textbf{Supplementary Table S5. Multi-$K$ overlap evaluation of predicted drugs against the curated migraine reference drug set} ($|\mathcal{R}| = 4{,}824$ normalized drugs). Results are shown for top-$N$ input genes ($N = 200$ and $N = 500$) under two evaluation universes. \textbf{ALL}: global background of 139{,}597 drugs; hypergeometric enrichment test is valid. \textbf{PREDICTED}: returned candidate pool only; FE values below 1.0 reflect the fact that the reference drug set exceeds the returned candidate pool in size, making standard hypergeometric enrichment interpretation inapplicable --- AUROC and AUPRC are the appropriate metrics in this frame. Within-set performance: for $N = 200$, AUROC\,$=$\,0.8004 and AUPRC\,$=$\,0.3528 (353 curated migraine drugs among 3{,}963 predicted drugs); for $N = 500$, AUROC\,$=$\,0.8152 and AUPRC\,$=$\,0.3311 (527 curated migraine drugs among 7{,}981 predicted drugs).}
\label{tab:S5_drug_multik}
\footnotesize
\setlength{\tabcolsep}{3pt}
\renewcommand{\arraystretch}{1.12}
\resizebox{\textwidth}{!}{%
\begin{tabular}{r l r r r r r r r r r}
\toprule
$N$ & \textbf{Universe} & $K$ & $|\mathcal{U}|$ & $|\mathcal{R}|$ & \textbf{Overlap} & \textbf{Prec@K} & \textbf{Rec@K} & \textbf{F1@K} & \textbf{Expected} & \textbf{FE} \\
\midrule
\multicolumn{11}{l}{\textbf{Top-$N = 200$ input genes}} \\
200 & ALL\_DRUGS & 20  & 139{,}597 & 4{,}824 & 5   & 0.250 & 0.001036 & 0.002064 & 0.691   & 7.235  \\
200 & ALL\_DRUGS & 50  & 139{,}597 & 4{,}824 & 20  & 0.400 & 0.004146 & 0.008207 & 1.728   & 11.575 \\
200 & ALL\_DRUGS & 100 & 139{,}597 & 4{,}824 & 51  & 0.510 & 0.010572 & 0.020715 & 3.456   & 14.758 \\
200 & ALL\_DRUGS & 200 & 139{,}597 & 4{,}824 & 89  & 0.445 & 0.018449 & 0.035430 & 6.911   & 12.877 \\
200 & ALL\_DRUGS & 500 & 139{,}597 & 4{,}824 & 205 & 0.410 & 0.042496 & 0.077010 & 17.278  & 11.865 \\
200 & PREDICTED  & 20  & 3{,}963   & 4{,}824 & 5   & 0.250 & 0.001036 & 0.002064 & 24.345  & 0.205  \\
200 & PREDICTED  & 50  & 3{,}963   & 4{,}824 & 20  & 0.400 & 0.004146 & 0.008207 & 60.863  & 0.329  \\
200 & PREDICTED  & 100 & 3{,}963   & 4{,}824 & 51  & 0.510 & 0.010572 & 0.020715 & 121.726 & 0.419  \\
200 & PREDICTED  & 200 & 3{,}963   & 4{,}824 & 89  & 0.445 & 0.018449 & 0.035430 & 243.452 & 0.366  \\
200 & PREDICTED  & 500 & 3{,}963   & 4{,}824 & 205 & 0.410 & 0.042496 & 0.077010 & 608.630 & 0.337  \\
\addlinespace
\multicolumn{11}{l}{\textbf{Top-$N = 500$ input genes}} \\
500 & ALL\_DRUGS & 20  & 139{,}597 & 4{,}824 & 8   & 0.400 & 0.001658 & 0.003303 & 0.691   & 11.575 \\
500 & ALL\_DRUGS & 50  & 139{,}597 & 4{,}824 & 20  & 0.400 & 0.004146 & 0.008207 & 1.728   & 11.575 \\
500 & ALL\_DRUGS & 100 & 139{,}597 & 4{,}824 & 53  & 0.530 & 0.010987 & 0.021527 & 3.456   & 15.337 \\
500 & ALL\_DRUGS & 200 & 139{,}597 & 4{,}824 & 93  & 0.465 & 0.019279 & 0.037022 & 6.911   & 13.456 \\
500 & ALL\_DRUGS & 500 & 139{,}597 & 4{,}824 & 213 & 0.426 & 0.044154 & 0.080015 & 17.278  & 12.328 \\
500 & PREDICTED  & 20  & 7{,}981   & 4{,}824 & 8   & 0.400 & 0.001658 & 0.003303 & 12.089  & 0.662  \\
500 & PREDICTED  & 50  & 7{,}981   & 4{,}824 & 20  & 0.400 & 0.004146 & 0.008207 & 30.222  & 0.662  \\
500 & PREDICTED  & 100 & 7{,}981   & 4{,}824 & 53  & 0.530 & 0.010987 & 0.021527 & 60.444  & 0.877  \\
500 & PREDICTED  & 200 & 7{,}981   & 4{,}824 & 93  & 0.465 & 0.019279 & 0.037022 & 120.887 & 0.769  \\
500 & PREDICTED  & 500 & 7{,}981   & 4{,}824 & 213 & 0.426 & 0.044154 & 0.080015 & 302.218 & 0.705  \\
\bottomrule
\end{tabular}%
}
\end{table}

\clearpage

\begin{table}[!ht]
\centering
\caption{\textbf{Supplementary Table S6. External contextual comparison of G2DR with the Open Targets migraine disease-target resource for migraine gene recovery.} The full G2DR analysis was evaluated across the complete ranked gene universe ($n = 9{,}305$ genes). The target-filtered G2DR analysis was restricted to genes represented in the Open Targets migraine target space ($n = 823$ genes). Overall recovery is shown against the curated migraine reference set ($M = 3{,}190$ genes), together with Top-$K$ precision before and after filtering. The substantially higher precision of standalone Open Targets (Top-50: 92.00\%) reflects its disease-curated, pre-filtered design rather than a direct performance comparison with G2DR. G2DR operates across a wider gene universe and recovered 725 curated migraine reference genes absent from Open Targets entirely, indicating that the two resources are complementary rather than directly competitive.}
\label{tab:S6_opentargets}
\footnotesize
\setlength{\tabcolsep}{4pt}
\renewcommand{\arraystretch}{1.12}
\begin{tabularx}{\textwidth}{>{\RaggedRight\arraybackslash}X >{\centering\arraybackslash}p{2.3cm} >{\centering\arraybackslash}p{2.8cm} >{\centering\arraybackslash}p{2.3cm}}
\toprule
\textbf{Metric} & \textbf{G2DR (all genes)} & \textbf{G2DR (target-filtered)} & \textbf{Open Targets} \\
\midrule
Ranked / returned genes & 9{,}305 & 823 & 2{,}376 \\
Reference genes recovered & 1{,}189 & 464 & 1{,}228 \\
Reference recovery (\%) & 37.27 & 14.55 & 38.50 \\
Shared with Open Targets & 823 & 823 & -- \\
Shared with Open Targets and reference & 464 & 464 & 464 \\
Reference genes recovered but absent from Open Targets & 725 & 0 & -- \\
Top-50 precision (\%) & 22.00 & 46.00 & 92.00 \\
Top-100 precision (\%) & 22.00 & 50.00 & 95.00 \\
Top-200 precision (\%) & 23.00 & 55.50 & 92.50 \\
Top-500 precision (\%) & 27.60 & 56.60 & 92.00 \\
\bottomrule
\end{tabularx}
\end{table}
\clearpage

\noindent\rule{\textwidth}{0.4pt}
\vspace{1em}
\begin{center}
\textit{End of Supplementary Material}
\end{center}